\documentclass{jpp-aas}
\usepackage[utf8]{inputenc}
\usepackage[T1]{fontenc}
\usepackage{amsmath}
\usepackage{amssymb}
\usepackage{graphicx}
\usepackage{subcaption}
\usepackage{commath}
\usepackage{upgreek}
\usepackage{appendix}
\usepackage{xcolor}
\usepackage{placeins}
\usepackage{hyperref}
\usepackage{todonotes}
\usepackage{multirow}
\usepackage{float}

\newcommand{\normvec}{\ensuremath{\boldsymbol{\hat{\mathrm{n}}}}}
\newcommand{\normpert}{\ensuremath{\delta\boldsymbol{\mathrm{x}}\cdot \normvec}}
\newcommand*\bs[1]{\mathop{}\!\boldsymbol{#1}}

\title{Exploration of the parameter space of quasisymmetric stellarator vacuum fields through adjoint optimisation}
\author{Richard Nies\aff{1,2}
  \corresp{\email{rnies@pppl.gov}},
  Elizabeth J. Paul\aff{3},
  Dario Panici\aff{4},
  Stuart R. Hudson\aff{2},
 \and Amitava Bhattacharjee\aff{1,2}}

\affiliation{
\aff{1}Department of Astrophysical Sciences, Princeton University, Princeton, NJ 08543, USA
\aff{2}Princeton Plasma Physics Laboratory, Princeton, NJ 08540, USA
\aff{3}Columbia University, New York, NY 10027, USA
\aff{4}Department of Mechanical Engineering, Princeton University, Princeton, NJ 08543, USA
}

\date{\today}

\begin{document}

\maketitle

\begin{abstract}
  Optimising stellarators for quasisymmetry leads to strongly reduced collisional transport and energetic particle losses compared to unoptimised configurations. Though stellarators with precise quasisymmetry have been obtained in the past, it remains unclear how broad the parameter space is where good quasisymmetry may be achieved. We study the range of aspect ratio and rotational transform values for which stellarators with excellent quasisymmetry on the boundary can be obtained. A large number of Fourier harmonics is included in the boundary representation, which is made computationally tractable by the use of adjoint methods to enable fast gradient-based optimisation, and by the direct optimisation of vacuum magnetic fields, which converge more robustly compared to solutions from magnetohydrostatics. Several novel configurations are presented, including stellarators with record levels of quasisymmetry on a surface; three field period quasiaxisymmetric stellarators with substantial magnetic shear, and compact quasisymmetric stellarators at low aspect ratios similar to tokamaks.
\end{abstract}

\section{Introduction}

Magnetic confinement fusion in toroidal geometry revolves around two main concepts, the axisymmetric tokamak and the three-dimensionally shaped stellarator \citep{spitzer_stellarator_1958}. The axisymmetry of the tokamak leads to greater engineering simplicity and good confinement of particles, at the cost of difficulty maintaining a stable steady-state plasma due to the need of a large plasma current. Stellarators avoid these issues by mostly relying on the external coils to provide the confining magnetic field, but they are generally plagued by poor neoclassical and energetic particle confinement at reactor-relevant low collisionality. However, the confinement may be returned to tokamak-like levels by making stellarators omnigenous \citep{hall_three-dimensional_1975} through careful optimisation of the three-dimensional geometry, as has been verified experimentally in the Wendelstein-7X device \citep{beidler_demonstration_2021}.

As a subset of omnigeneity, quasisymmetry \citep{nuhrenberg_quasi-helically_1988} (QS) recovers tokamak-like guiding-centre dynamics through a symmetry in the magnetic field strength $B=B(s, M\theta_B - N \zeta_B)$, with the flux surface label $s$, the Boozer poloidal and toroidal angles $\theta_B$ and $\zeta_B$, and the QS helicities $M$ and $N$. Depending on the helicities, QS configurations may be either quasiaxisymmetric (QA, $N=0$), quasihelical (QH, $M, N \neq 0$) or quasipoloidal (QP, $M=0$). Several QS stellarators have been designed, e.g. the quasiaxisymmetric NCSX \citep{zarnstorff_physics_2001} and MUSE \citep{qian_design_2023}, and the quasihelical HSX \citep{anderson_helically_1995}, the latter two of which were constructed. Recently, configurations with record levels of quasisymmetry have been obtained in optimisations of vacuum fields \citep{landreman_magnetic_2022} and finite-pressure equilibria \citep{landreman_optimization_2022}.

Most quasisymmetry optimisations to date have relied on derivative-free algorithms \citep{drevlak_estell_2013, bader_stellarator_2019, henneberg_properties_2019, henneberg_improving_2020}, or on gradient-based algorithms using finite-differences to evaluate the gradient \citep{landreman_magnetic_2022, landreman_optimization_2022}. Due to the large parameter space of the stellarator boundary representation, these schemes are computationally expensive, which has motivated the introduction of automatic differentiation techniques \citep{dudt_desc_2023} and adjoint methods \citep{paul_gradient-based_2021}. Moreover, the accuracy of gradients evaluated through finite-differences can be sensitive to the step size, requiring careful scans in the numerical parameters to ensure success in the optimisation, e.g. avoiding artificial local minima.

Though many configurations with good quasisymmetry have been obtained in the past, the degree of compatibility of QS with other objectives remains unclear\footnote{A notable exception is the recent study by \citet{buller2024family} exploring the compatiblity of QA with varying rotational transform values for $N_\mathrm{fp}=2$ (two field-period) stellarators.} Furthermore, slower optimisation algorithms limit the amount of shaping that can be taken advantage of in the optimisation.  To remedy these gaps, we here perform quasisymmetry optimisation of stellarators using adjoint methods. The fast optimisation facilitates the exploration of a large parameter space in the stellarator boundary shape, and the compatibility of QS with aspect ratio and rotational transform targets. We optimise vacuum magnetic fields, instead of considering the vacuum limit (zero plasma current and pressure) of magnetohydrostatics (MHS) with the imposition of nested flux surfaces, as computed e.g. by the codes VMEC \citep{hirshman_three-dimensional_1986} and DESC \citep{dudt_desc_2020}. The vacuum magnetic field solution converges more robustly than the solution from MHS due to the linearity of Laplace's equation, allowing us to go to high resolutions reliably.

In \S\ref{sec:methods}, our objective function and optimisation methods are introduced. In \S\ref{sec:aspect_ratio_scan}~and~\ref{sec:iota_scan}, we then present the obtained quasisymmetric configurations with varying aspect ratio and rotational transform, respectively. In \S\ref{sec:Auxiliary_opt}, we show how the optimised configurations may be further refined, e.g. by optimising for quasisymmetry in the full volume (\S\ref{sec:postproc_QA_volume_opt}), or by improving the nestedness of flux-surfaces in a QA with substantial magnetic shear (\S\ref{sec:postproc_integrability_opt}). Finally, we present our conclusions in \S\ref{sec:conclusions}.

\section{Methods} \label{sec:methods}

The objective function $f$ considered in the optimisations presented herein contains three targets,
\begin{equation}\label{eq:f_tot}
    f = f_\mathrm{QS}^\star + w_\iota f_\iota + w_A f_A,
\end{equation}
with weights $w_\iota$ and $w_A$. The objective includes a target for QS on the boundary $f_\mathrm{QS}^\star$, defined in \eqref{eq:def_fQS_star}, another for the edge rotational transform $\iota_e$,
\begin{equation}
    f_\iota = \frac{1}{2}\left( \iota_e - \iota_T \right)^2,
\end{equation}
with prescribed $\iota_T$, and finally an aspect ratio objective,
\begin{equation}
    f_A = \frac{1}{2}\left( A - A_T \right)^2, \label{eq:f_aspectratio}
\end{equation}
with prescribed $A_T$. The aspect ratio definition \citep[see e.g.][]{landreman_constructing_2019} follows that employed in the VMEC code \citep{hirshman_three-dimensional_1986}, 
\begin{equation}
    A = \frac{R_\mathrm{maj}}{a_\mathrm{min}}, \quad \text{with} \quad R_\mathrm{maj} = \frac{\mathcal{V}}{2\pi^2 a_\mathrm{min}^2}, \quad a_\mathrm{min} = \sqrt{\frac{\overline{S}}{\pi}}, \quad \overline{S} = \frac{1}{2\pi}\int_0^{2\pi}\mathrm{d}\zeta\; S(\zeta), \label{eq:def_aspectratio}
\end{equation}
with $S(\zeta)$ the cross-sectional area of the boundary at fixed toroidal angle $\zeta$, and $\mathcal{V}$ the volume enclosed by the boundary.

The optimisation is performed over a set of harmonics $R_{mn}$ and $Z_{mn}$ describing the boundary, here assuming stellarator symmetry,
\begin{subequations} \label{eqs:boundary_rep}
\begin{align}
    R(\theta, \zeta) = \sum_{m, n}  R_{mn} \cos\left( m\theta - n N_\mathrm{fp} \zeta \right), \label{eq:Rmn} \\
    Z(\theta, \zeta) = \sum_{m, n} Z_{mn} \sin\left( m\theta - n N_\mathrm{fp} \zeta \right), \label{eq:Zmn}
\end{align}
\end{subequations}
with $N_\mathrm{fp}$ the number of field periods of the device, leading to a discrete symmetry under the transformation $\zeta \rightarrow \zeta + 2\pi/N_\mathrm{fp}$. In the optimisation, the series in \eqref{eqs:boundary_rep} are truncated at some chosen $m_\mathrm{max}$ and $n_\mathrm{max}$. Furthermore, $R_{00}=1$ is kept constant to set the length scale of the problem, as the objective \eqref{eq:f_tot} is invariant under a uniform rescaling of all $R_{mn}$ and $Z_{mn}$ coefficients.

The derivatives of the quasisymmetry and rotational transform targets with respect to the boundary coefficients is obtained using adjoint methods. The corresponding adjoint equations and shape gradient were originally derived in \cite{nies_adjoint_2022}, and are here slightly modified to make the quasisymmetry objective dimensionless, as shown in Appendix~\ref{app:normalised_qs_obj}. The derivative of the aspect ratio target \eqref{eq:f_aspectratio} is derived in Appendix~\ref{app:aspectratio_sg}.

Each optimisation begins with small values of $m_\mathrm{max}$ and $n_\mathrm{max}$ before gradually increasing them in optimisation `stages', letting the optimisation run in each such stage until progress has halted. The Fourier space resolution of the solutions to the Laplace equation for the vacuum magnetic field, to the straight-field line equation, and to their adjoint equations, is always higher than the boundary resolution and is also increased at each stage. We found it generally optimal to start the optimisation with a single poloidal harmonic $m_\mathrm{max}=1$, but with a larger number of toroidal harmonics, e.g. $n_\mathrm{max} = 3$. This may be due to the connection between the axis shape and quasisymmetry \citep{rodriguez_phases_2022}. Owing to the use of adjoint methods, a large parameter space in the boundary representation \eqref{eqs:boundary_rep} may be explored, such that our optimisations typically involve a dozen stages or more, incrementally going up to $m_\mathrm{max}\sim n_\mathrm{max} \sim 10-20$. Despite the high resolution, a typical optimisation requires only $\mathcal{O}(10^2)$ CPU-hours, with $\mathcal{O}(10^3)$ iterations of the optimiser over the multiple stages.

All optimisations are performed using the BFGS algorithm \citep{fletcher_practical_1987} implemented in the \texttt{scipy} package \citep{virtanen_scipy_2020}, with the initial boundary taken to be that of a simple rotating ellipse. The Laplace equation for the vacuum magnetic field, as well as the corresponding adjoint problem \citep{nies_adjoint_2022}, are solved with the SPEC code \citep{hudson_computation_2012, qu_coordinate_2020}. The rotational transform on the boundary is obtained by solving the straight-field line equation $\bs{B}\cdot\nabla\alpha = 0$, with the field line label $\alpha = \theta - \iota \zeta + \lambda(\theta, \zeta)$, where $\lambda$ is a single-valued function of the angular coordinates.

After an optimised vacuum field boundary has been obtained, it may then be input to VMEC \citep{hirshman_three-dimensional_1986} or DESC \citep{dudt_desc_2020, panici_desc_2023, conlin_desc_2023, dudt_desc_2023}, solving MHD force balance while setting the pressure and current profiles to vanish. This procedure yields a good approximation to the vacuum field, provided the flux surfaces are nested. This is generally the case for the quasihelical configurations, as well as the quasiaxisymmetric configurations with $N_\mathrm{fp}=2$, but not for those with $N_\mathrm{fp}=3$ due to their larger magnetic shear, see \S\ref{sec:aspect_ratio_scan}~and~\S\ref{sec:iota_scan}. The DESC solution is used only for post-processing, to evaluate volume properties and the magnetic axis. The VMEC code is used in conjunction with the SIMSOPT \citep{landreman_simsopt_2021} optimisation suite in \S\ref{sec:Auxiliary_opt}, to further optimise two configurations.

Although $f_\mathrm{QS}^\star$ from \eqref{eq:def_fQS_star} is used in the optimisation, for post-processing the degree of QS is also evaluated using the infinity-norm of the symmetry-breaking modes
\begin{equation} \label{eq:symmetry_breaking_Bmn}
    \abs{B_{mn}}_\infty \equiv \max_{n/m\neq N/M} B_{mn},
\end{equation}
which may be evaluated on the boundary from either the SPEC or the DESC solution. Here, $B_{mn}$ are the coefficients of the field strength when Fourier expanded in the Boozer angles $\theta_B$ and $\zeta_B$, i.e. $B = \sum B_{mn} \cos(m\theta_B - n N_\mathrm{fp} \zeta_B)$. For the SPEC vacuum field solution, the Boozer coordinate transformation required to evaluate \eqref{eq:symmetry_breaking_Bmn} follows the procedure presented in Appendix~\ref{app:Boozer_transf}. From the DESC solution, we further define a metric of QS in the volume as
\begin{equation}
    \left\langle \frac{\abs{B_{mn}}_\infty}{B_{00}} \right\rangle = \int_0^1\mathrm{d}s\; \frac{\abs{B_{mn}}_\infty}{B_{00}},
\end{equation}
where $s$ is a normalised flux coordinate ranging from $s=0$ on the magnetic axis to $s=1$ on the boundary. We note that although perfect quasisymmetry would result in both $f_\mathrm{QS}^\star=0$ and $\abs{B_{mn}}_\infty = 0$, the two measures of QS employed in this study do not perfectly correlate as QS is approached \citep{rodriguez_measures_2022}, such that larger $\abs{B_{mn}}_\infty$ values than expected might be obtained in the configurations optimised for low $f_\mathrm{QS}^\star$. Moreover, we observe differences in the level of QS on the boundary given by the DESC and vacuum field solutions, see \S\ref{sec:aspect_ratio_scan}~and~\S\ref{sec:iota_scan}. These may be due to the different algorithms employed, or small inaccuracies in the equilibrium solutions becoming important as very small QS deviations are reached.

\section{Aspect ratio scan} \label{sec:aspect_ratio_scan}

We first investigate how the choice of aspect ratio $A$ affects the quasisymmetry optimisation. We consider QA configurations with $N_\mathrm{fp}=2$ and $N_\mathrm{fp}=3$, with a boundary rotational transform target $\iota_T = 0.42$, as had been used for the precise QA configuration obtained by \citet{landreman_magnetic_2022}. Optimisations for QA at $N_\mathrm{fp}=1$ and $N_\mathrm{fp}=4$ generally performed poorly, consistent with previous studies \citep{Landreman_2022}, and are thus not shown here. For the QH configurations, $N_\mathrm{fp}=3$ and $N_\mathrm{fp}=4$ are considered, with rotational transform targets of $\iota_T \approx 1.3$. Though it is crucial only for the QA to prescribe a rotational transform target, as the optimisation would otherwise tend towards an axisymmetric solution with $\iota=0$, we also included an $\iota$ target for the QH to make the comparison between different aspect ratio values more meaningful. We do not include here QH configurations with either $N_\mathrm{fp}= 2$, due to their larger deviations from QS in our optimisation, or with $N_\mathrm{fp}\geq 5$, as these could only be optimised at larger aspect ratio values.

For the QA configuration with $N_\mathrm{fp}=2$ and $A=6$, the optimisation was pushed to very high resolutions to reach record levels of QS on the boundary. This optimisation was performed as a proof of principle that the QS error on the boundary could be reduced substantially, though a large number of boundary harmonics is required in practice, and an increased computational cost is incurred. We note that reducing the QS error from $\abs{B_{mn}}_\infty/B_{00} \sim 10^{-5}$ to $\abs{B_{mn}}_\infty/B_{00}\sim 10^{-10}$ does not substantially change the boundary shape, or the average QS error in the volume.

\begin{figure}
\captionsetup[subfigure]{labelformat=empty}
     \centering
     \begin{subfigure}[t]{0.24\textwidth}
         \centering
         \includegraphics[width=\textwidth, trim = 0cm 1.5cm 0cm 0cm, clip]{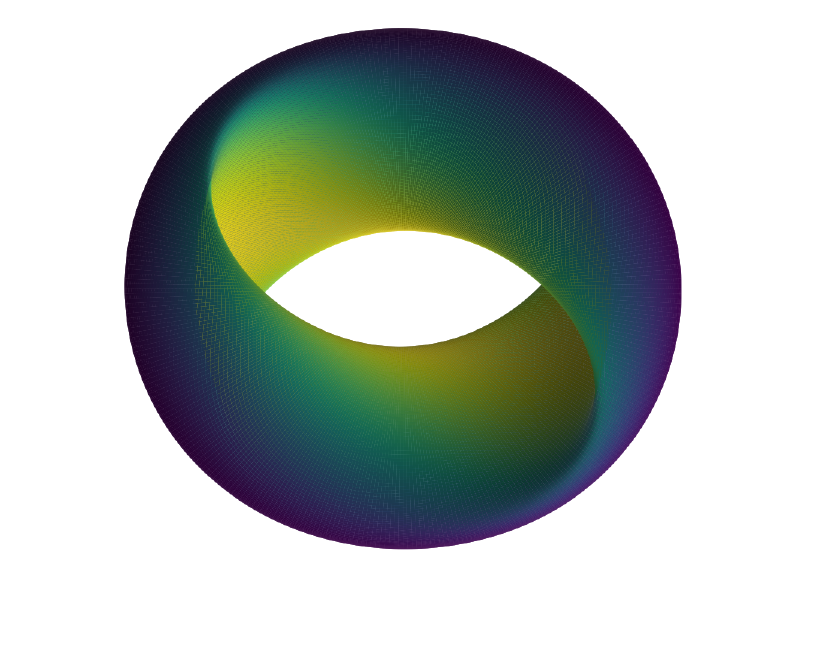}
         \label{fig:QA-Nfp2-AR-2.6-modB-top}
     \end{subfigure}
     \hfill
     \begin{subfigure}[t]{0.24\textwidth}
         \centering
         \includegraphics[width=\textwidth, trim = 0cm 1.5cm 0cm 0cm, clip]{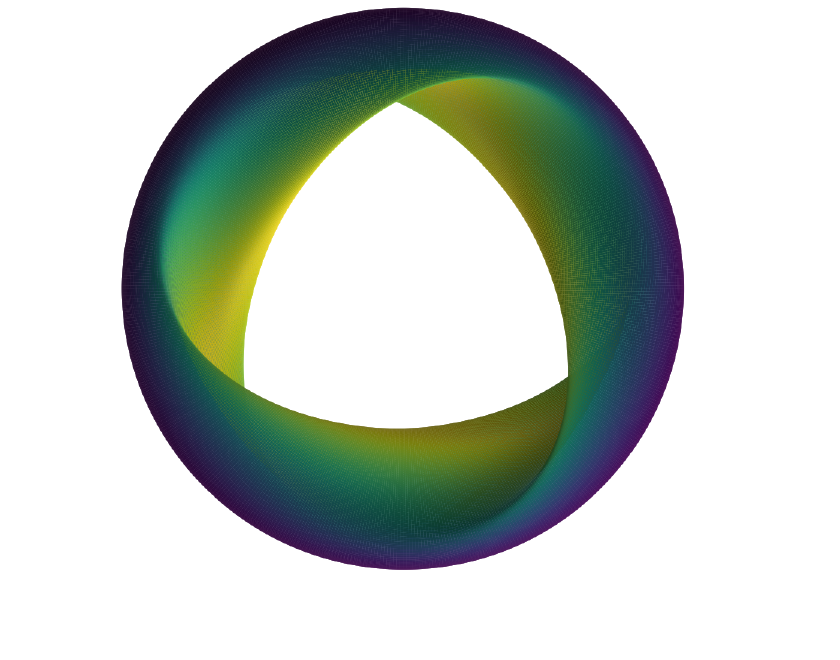}
         \label{fig:QA-Nfp3-AR-4.5-modB-top}
     \end{subfigure}
     \hfill
      \begin{subfigure}[t]{0.24\textwidth}
         \centering
         \includegraphics[width=\textwidth, trim = 0cm 1.5cm 0cm 0cm, clip]{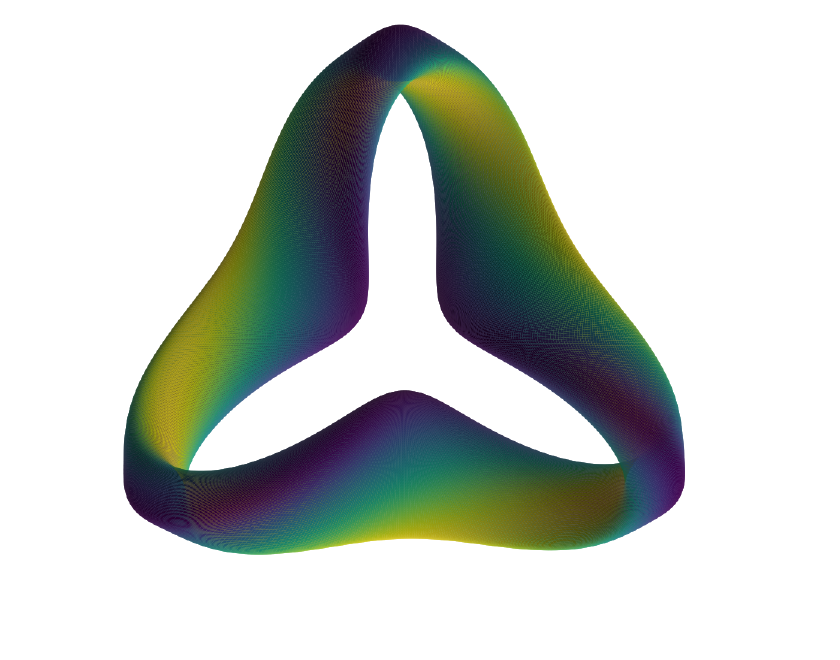}
         \label{fig:compact-QH-Nfp3-top}
     \end{subfigure}
     \hfill
     \begin{subfigure}[t]{0.24\textwidth}
         \centering
         \includegraphics[width=\textwidth, trim = 0cm 1.5cm 0cm 0cm, clip]{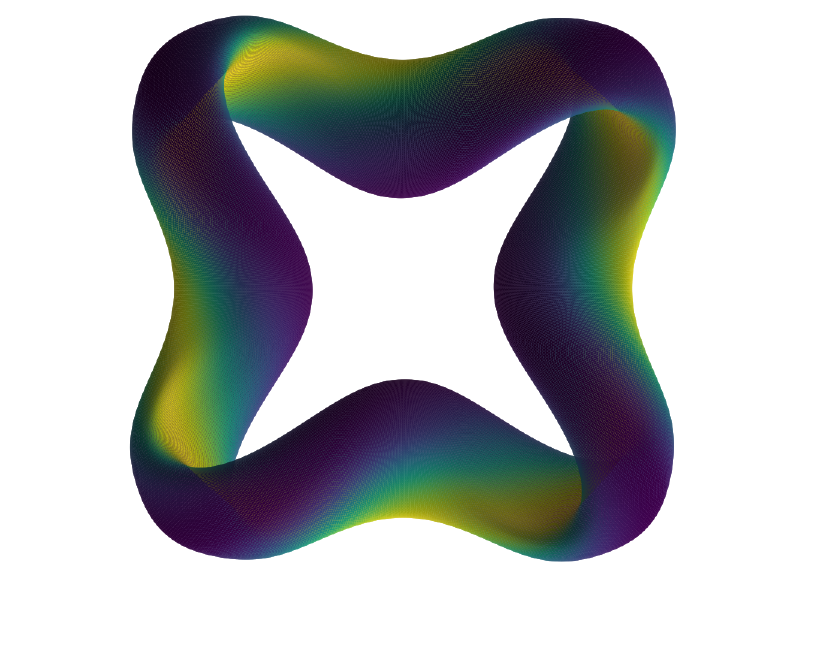}
         \label{fig:compact-QH-Nfp4-top}
     \end{subfigure}
     \\
     % \vspace{-7.3mm}
     
    \begin{subfigure}[t]{0.24\textwidth}
         \centering
         \includegraphics[width=\textwidth, trim = 0cm 3.5cm 0cm 2.2cm, clip]{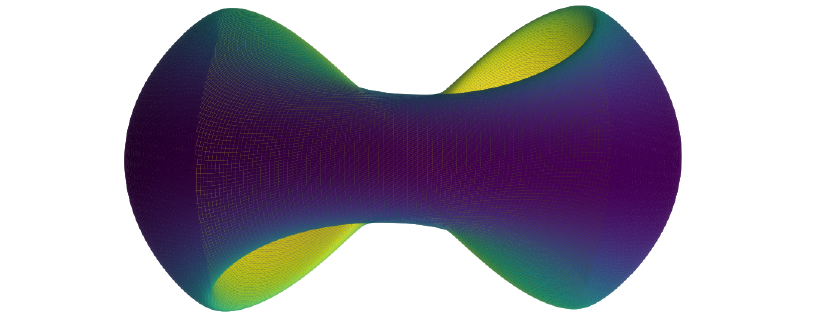}
         \label{fig:QA-Nfp2-AR-2.6-modB-side}
     \end{subfigure}
    \hfill     
    \begin{subfigure}[t]{0.24\textwidth}
         \centering
         \includegraphics[width=\textwidth, trim = 0cm 3.5cm 0cm 2.2cm, clip]{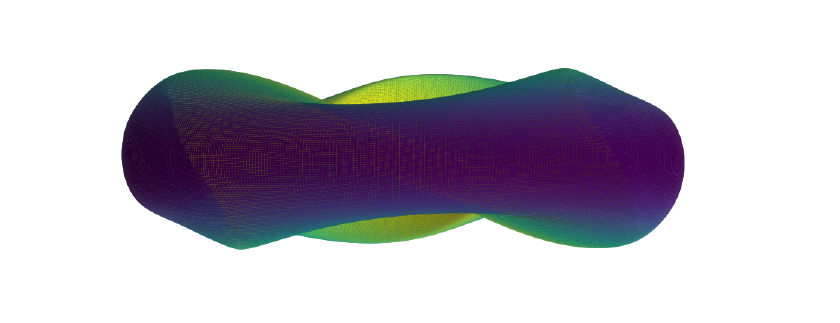}
         \label{fig:QA-Nfp3-AR-4.5-modB-side}
     \end{subfigure}
     \hfill
     \begin{subfigure}[t]{0.24\textwidth}
         \centering
         \includegraphics[width=\textwidth, trim = 0cm 3.5cm 0cm 2.2cm, clip]{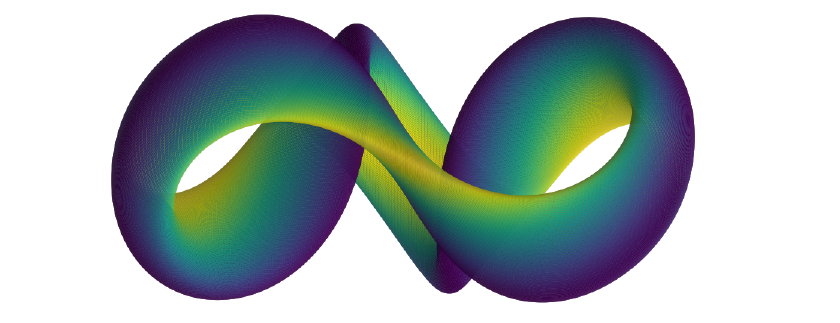}
         \label{fig:compact-QH-Nfp3-side}
     \end{subfigure}
     \hfill
     \begin{subfigure}[t]{0.24\textwidth}
         \centering
         \includegraphics[width=\textwidth, trim = 0cm 3.5cm 0cm 2.2cm, clip]{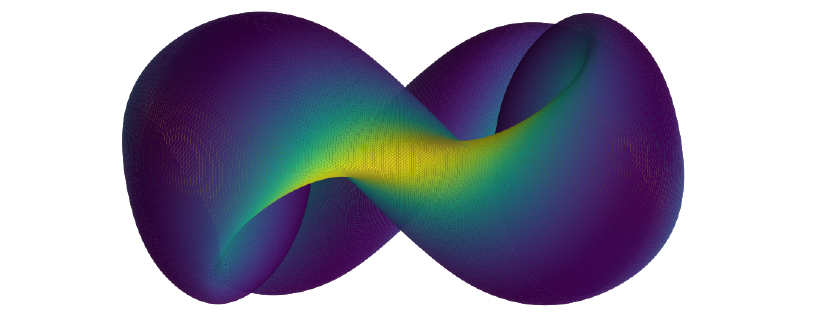}
         \label{fig:compact-QH-Nfp4-side}
     \end{subfigure}
     \\
     % \vspace{-12.2mm}
     
    \begin{subfigure}[t]{0.24\textwidth}
         \centering
         \includegraphics[width=\textwidth, trim = 0cm 0cm 0cm 1.4cm, clip]{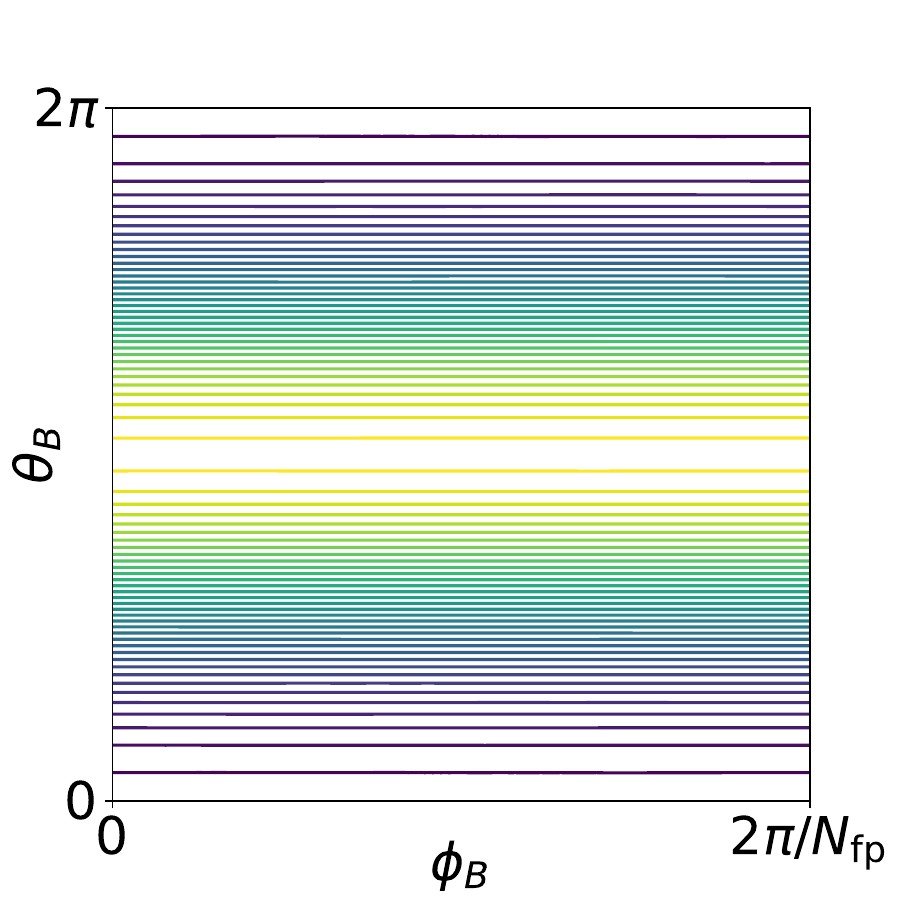}
         \caption{$A=2.6$ $N_\mathrm{fp}=2$ QA}
     \end{subfigure}
     \hfill
     \begin{subfigure}[t]{0.24\textwidth}
         \centering
         \includegraphics[width=\textwidth, trim = 0cm 0cm 0cm 1.4cm, clip]{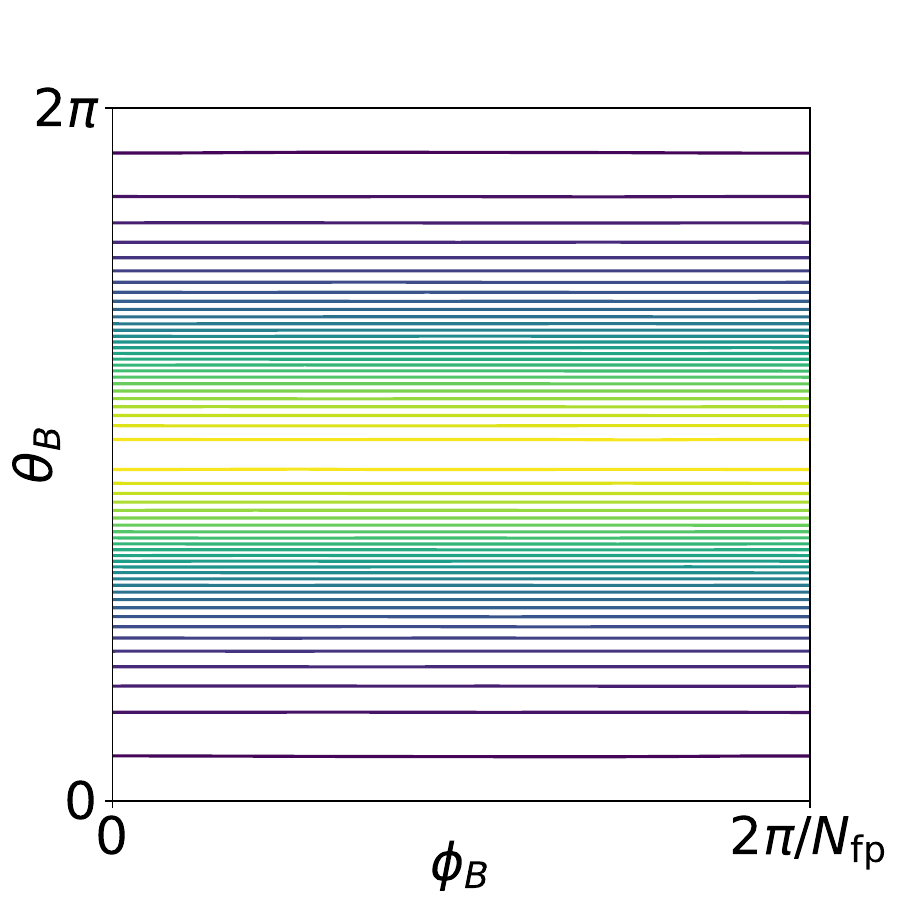}
         \caption{$A = 4.5$ $N_\mathrm{fp}=3$ QA}
     \end{subfigure}
    \hfill     
     \begin{subfigure}[t]{0.24\textwidth}
         \centering
         \includegraphics[width=\textwidth, trim = 0cm 0cm 0cm 1.4cm, clip]{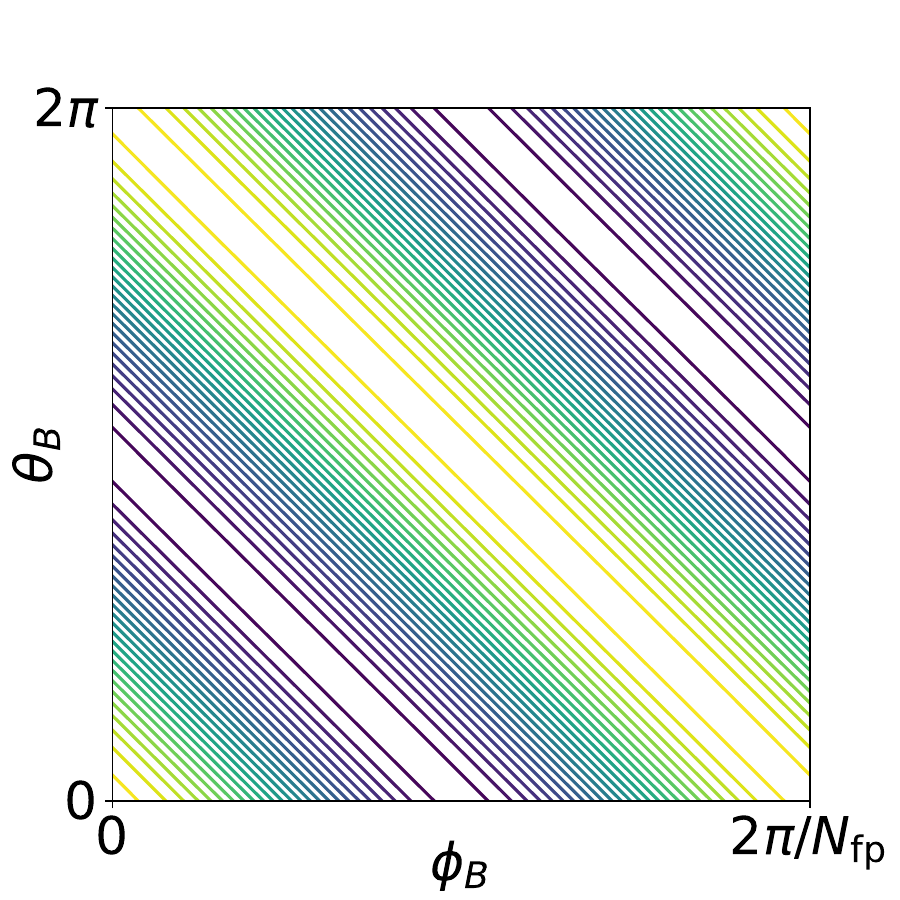}
         \caption{$A=3.6$ $N_\mathrm{fp}=3$ QH}
         \label{fig:compact-QH-Nfp3-Boozer}
     \end{subfigure}
     \hfill
    \begin{subfigure}[t]{0.24\textwidth}
         \centering
         \includegraphics[width=\textwidth, trim = 0cm 0cm 0cm 1.4cm, clip]{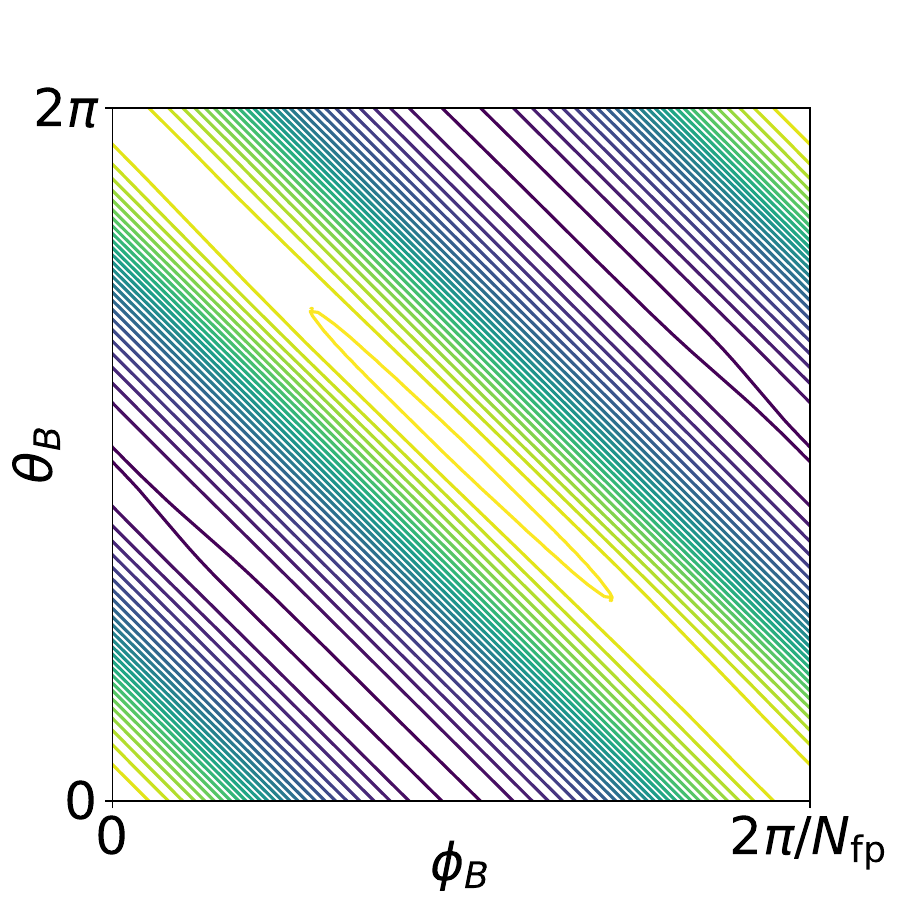}
         \caption{$A=3$ $N_\mathrm{fp}=4$ QH}
         \label{fig:compact-QH-Nfp4-Boozer}
     \end{subfigure}
        \caption{Contours of the magnetic field strength $B$ on the boundary of quasisymmetric stellarators at small aspect ratio $A$: top view (first row), side view (second row) and in Boozer coordinates (third row).}
         \label{fig:optimisation_highlights_AR_scan}
\end{figure}

The contours of the magnetic field strength on the boundary of the configurations with the lowest aspect ratio obtained for each case are displayed in figure~\ref{fig:optimisation_highlights_AR_scan}. Configurations with good QA (i.e. those for which the boundary magnetic field strength contours look approximately straight in Boozer coordinates) could be found down to $A=2.6$ and $A=4.5$ for $N_\mathrm{fp}=2$ and $N_\mathrm{fp}=3$, respectively. For the QH configurations, the aspect ratio could be reduced to $A=3.6$ and $A=3$ for $N_\mathrm{fp} = 3$ and $N_\mathrm{fp}=4$, respectively. As attested by the contours in Boozer coordinates (which are perfectly straight in the limit of exact QS), these configurations have excellent QS on the boundary. Only the low aspect ratio $N_\mathrm{fp}=4$ QH configuration has visible deviations of the contours from straight lines. For the QA configurations, the three-dimensional shaping is visibly strongest on the inboard side, in agreement with the analytical results of \cite{plunk_quasi-axisymmetric_2018} for low aspect ratio QA stellarators close to axisymmetry.

\begin{figure}
     \centering
     \begin{subfigure}[t]{0.49\textwidth}
         \centering
            \includegraphics[width=\textwidth, trim = 0.4cm 0cm 0.4cm 0cm, clip]{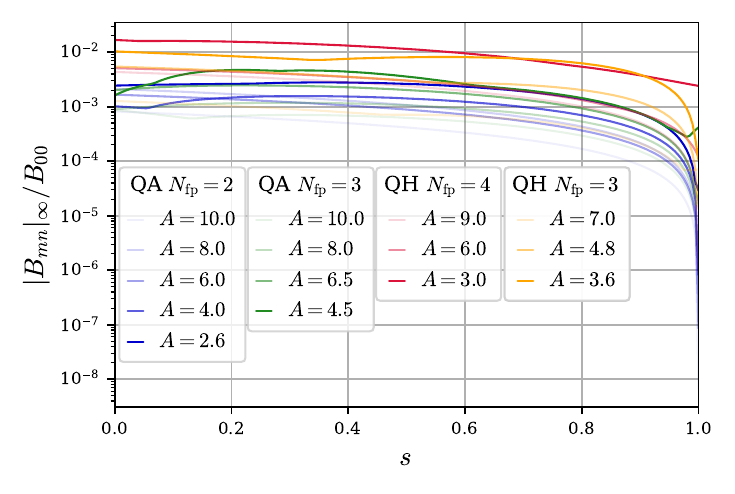}
         % \caption{Quasisymmetry error}
         % \label{fig:QA-Nfp2-iota-scan_axis_side-bmnc}
     \end{subfigure}
     \hfill
     \begin{subfigure}[t]{0.49\textwidth}
         \centering
         \includegraphics[width=\textwidth, trim = 0.4cm 0cm 0.4cm 0cm, clip]{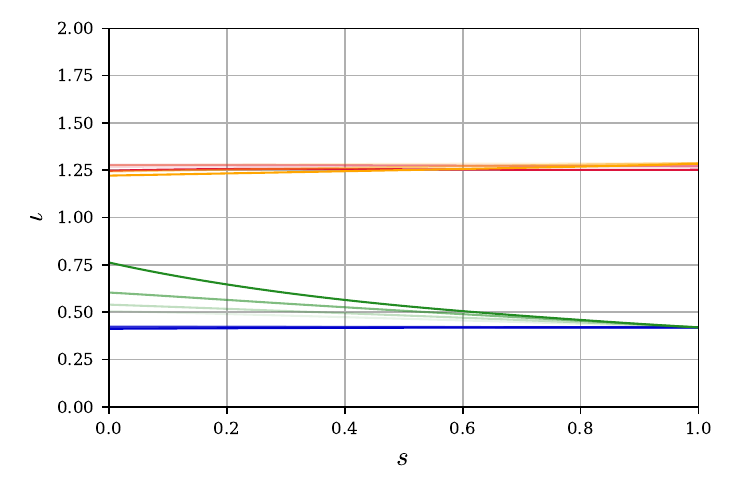}
         % \caption{Rotational transform}
         % \label{fig:QA-Nfp2-iota-scan_axis_iota}
     \end{subfigure}
        \caption{Volume properties of stellarators optimised for quasisymmetry on the boundary, with varying aspect ratio $A$. The QS error (left plot) and rotational transform (right plot) profiles were obtained using the DESC code. The QS error reaches record levels on the boundary ($s=1$), and generally remains relatively low in the core. The rotational transform profiles are flat for all configurations except for $N_\mathrm{fp}=3$ QA. }
        \label{fig:summary_AR_scan}
\end{figure}

The quasisymmetry error and rotational transform profiles for all configurations in the aspect ratio scan are shown in figure~\ref{fig:summary_AR_scan} as a function of the normalised flux $s$. Further properties of the configurations are listed in table~\ref{tab:aspectratio}. As expected, the quasisymmetry is best on the boundary, down to record normalised values of $\abs{B_{mn}}_\infty/B_{00} \sim 10^{-8}$ in the DESC solution, and $\abs{B_{mn}}_\infty/B_{00} \sim 10^{-10}$ in the SPEC solution. The quasisymmetry error also remains low throughout the volume in most cases, $\abs{B_{mn}}_\infty/B_{00} \lesssim 10^{-2}$ at small aspect ratio, and $\abs{B_{mn}}_\infty/B_{00} \lesssim 10^{-3}$ at large aspect ratio. The good QS levels throughout the volume suggest that, in practice, the optimisation of QS on a boundary is achieved by approximating a globally QS configuration, see \S\ref{sec:postproc_QA_volume_opt}. 

The QA $N_\mathrm{fp}=2$ and $N_\mathrm{fp}=3,4$ QH configurations all have small magnetic shear, as attested by the flatness of the rotational transform profiles, in accordance with previous QS optimisations of vacuum fields \citep{landreman_magnetic_2022}. However, the $N_\mathrm{fp}=3$ QA stellarators attain substantial magnetic shear, with $\iota$ varying from $0.42$ on the boundary up to $0.75$ on axis. In those cases, the magnetic shear $\mathrm{d}\iota/\mathrm{d}\psi$ is positive (tokamak-like) and approximately constant, such that $\Delta\iota = \iota(s=1)-\iota(s=0) \sim \mathrm{d}\iota/\mathrm{d}\psi \;\Delta \psi \propto 1/A^2$, as the change in flux $\Delta\psi \sim B r^2 \sim (B R^2) / A^2$.

The substantial magnetic shear of the $N_\mathrm{fp}=3$ QA configurations means the rotational transform generally crosses low-order rational values in the core, leading to island chains and chaotic regions. In those cases, the DESC solution may not be trustworthy, as it assumes nested flux surfaces a priori. However, the integrability may be improved in an auxiliary optimisation, as shown in \S\ref{sec:postproc_integrability_opt}. We note that large chaotic regions are found for the lowest aspect ratio $N_\mathrm{fp}=3$ QA with $A=4.5$. One may hypothesise that the aspect ratio is here limited by the chaotic region increasing in size until it encounters the boundary.

\begin{table}
    \centering
    \begin{tabular}{c|c|c|c|c|c|c|c}
        & $A$ & $f_\mathrm{QS}^\star$ & $\left(\frac{\abs{B_\mathrm{mn}}_\infty}{B_{00}}\right)_{e,\mathrm{SPEC}}$ & $\left(\frac{\abs{B_\mathrm{mn}}_\infty}{B_{00}}\right)_{e,\mathrm{DESC}}$ & $\left\langle\frac{\abs{B_\mathrm{mn}}_\infty}{B_{00}}\right\rangle$ & $\Delta\iota$ & $\kappa_\mathrm{min} R_{00}$ \\\hline
\multirow{5}{6em}{QA $N_\mathrm{fp}=2$} & 10.0 & $5\cdot10^{-8}$ & $3\cdot10^{-9}$ & $7\cdot10^{-9}$ & $4\cdot10^{-4}$ & $-1\cdot10^{-3}$ & 0.50\\
                                        & 8.0  & $4\cdot10^{-8}$ & $2\cdot10^{-9}$ & $4\cdot10^{-7}$ & $1\cdot10^{-3}$ & $-6\cdot10^{-3}$ & 0.51\\
                                        & 6.0  & $2\cdot10^{-9}$ &$2\cdot10^{-10}$ & $9\cdot10^{-8}$ & $9\cdot10^{-4}$ & $-4\cdot10^{-3}$ & 0.49\\ % machine precision QA
                                        % & 6.0  & $5\cdot10^{-7}$ & $2\cdot10^{-8}$ &  &  & \\ % from scan
                                        & 4.0  & $1\cdot10^{-5}$ & $5\cdot10^{-7}$ & $8\cdot10^{-7}$ & $1\cdot10^{-3}$ & $-2\cdot10^{-4}$ & 0.48\\
                                        & 2.6  & $3\cdot10^{-4}$ & $2\cdot10^{-5}$ & $3\cdot10^{-5}$ & $2\cdot10^{-3}$ & $+7\cdot10^{-3}$ & 0.43\\\hline
\multirow{4}{6em}{QA $N_\mathrm{fp}=3$} & 10.0 & $1\cdot10^{-5}$ & $3\cdot10^{-7}$ & $2\cdot10^{-6}$ & $5\cdot10^{-4}$ & $-0.088$         & 0.54\\
                                        & 8.0  & $3\cdot10^{-5}$ & $7\cdot10^{-7}$ & $2\cdot10^{-5}$ & $9\cdot10^{-4}$ & $-0.120$         & 0.52\\
                                        & 6.5  & $2\cdot10^{-5}$ & $1\cdot10^{-6}$ & $2\cdot10^{-5}$ & $2\cdot10^{-3}$ & $-0.184$         & 0.51\\
                                        & 4.5  & $3\cdot10^{-3}$ & $6\cdot10^{-5}$ & $4\cdot10^{-4}$ & $3\cdot10^{-3}$ & $-0.343$         & 0.51\\\hline
\multirow{3}{6em}{QH $N_\mathrm{fp}=4$} & 9.0  & $2\cdot10^{-5}$ & $9\cdot10^{-7}$ & $9\cdot10^{-7}$ & $2\cdot10^{-3}$ & $+2\cdot10^{-3}$ & 1.12\\
                                        & 6.0  & $5\cdot10^{-4}$ & $2\cdot10^{-5}$ & $1\cdot10^{-4}$ & $3\cdot10^{-3}$ & $-8\cdot10^{-3}$ & 1.32\\
                                        & 3.0  & $3\cdot10^{-2}$ & $2\cdot10^{-3}$ & $2\cdot10^{-3}$ & $1\cdot10^{-2}$ & $+6\cdot10^{-3}$ & 1.12\\\hline
\multirow{3}{6em}{QH $N_\mathrm{fp}=3$} & 7.0  & $2\cdot10^{-3}$ & $7\cdot10^{-5}$ & $7\cdot10^{-5}$ & $7\cdot10^{-4}$ & $+0.011$         & 0.67\\
                                        & 4.8  & $2\cdot10^{-4}$ & $1\cdot10^{-5}$ & $1\cdot10^{-5}$ & $3\cdot10^{-3}$ & $+0.044$         & 0.66\\
                                        & 3.6  & $2\cdot10^{-3}$ & $1\cdot10^{-4}$ & $1\cdot10^{-4}$ & $7\cdot10^{-3}$ & $+0.063$         & 0.62
    \end{tabular}
    \caption{Properties of stellarators optimised for quasisymmetry at varying aspect ratio $A$: QS figure of merit $f_\mathrm{QS}^\star$, maximum symmetry-breaking mode on boundary from vacuum solution (SPEC) and MHS solution (DESC), volume-averaged QS error in MHS solution, difference between $\iota$ on axis and at the edge $\Delta \iota = \iota(s=1)-\iota(s=0)$, and minimum value of axis curvature $\kappa_\mathrm{min}$.}
    \label{tab:aspectratio}
\end{table}

%%%%%%%%%%%%%%%%%%%%%%%%%%%%%%%%%%%%%%%%%%%%%%%%%%%%%%%
\section{Rotational transform scan} \label{sec:iota_scan}

We now vary the boundary rotational transform target $\iota_T$ in the quasisymmetry optimisation. We again consider $N_\mathrm{fp}=2$ and $N_\mathrm{fp}=3$ QA, with a fixed aspect ratio of $A=6$. We also optimise $N_\mathrm{fp}=4$ QH stellators at a higher aspect ratio $A=8$. The aspect ratio targets were chosen to be identical to the precise $N_\mathrm{fp}=2$ QA and $N_\mathrm{fp}=4$ QH configurations of \cite{landreman_magnetic_2022}. Good quasisymmetry could be obtained for rotational transform values up to $\iota_e \sim 0.7$ and $\iota_e \sim 0.8$ for the $N_\mathrm{fp}=2$ and $N_\mathrm{fp}=3$ QA configurations, respectively, and in the range $\iota_e \sim 1$ to $\iota_e \sim 2$ for the $N_\mathrm{fp}=4$ QH. 

\begin{figure}
\captionsetup[subfigure]{labelformat=empty}
     \centering
     \begin{subfigure}[t]{0.24\textwidth}
         \centering
         \includegraphics[width=\textwidth, trim = 0cm 1.5cm 0cm 0cm, clip]{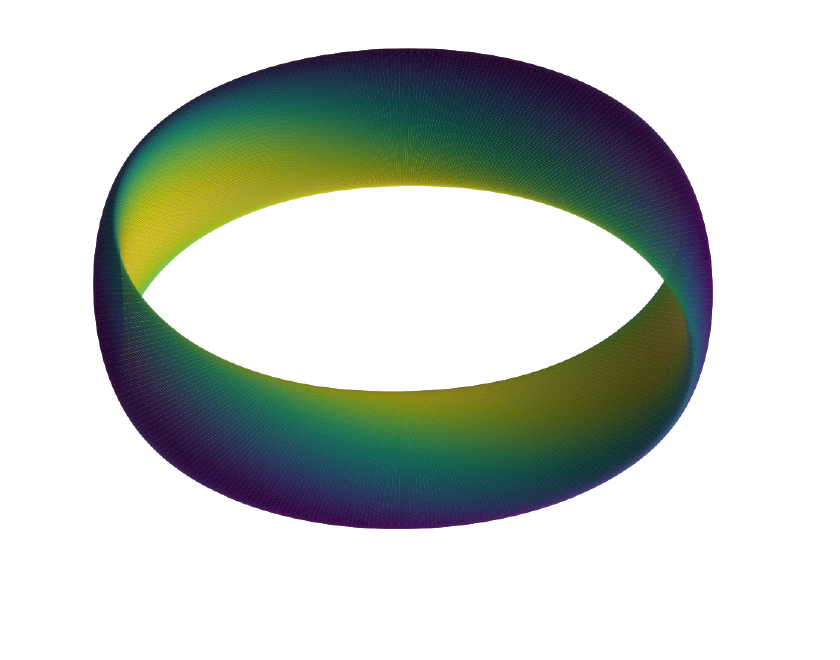}
         \label{fig:QA-Nfp2-iota-0.65-modB-top}
     \end{subfigure}
     \hfill
     \begin{subfigure}[t]{0.24\textwidth}
         \centering
         \includegraphics[width=\textwidth, trim = 0cm 1.5cm 0cm 0cm, clip]{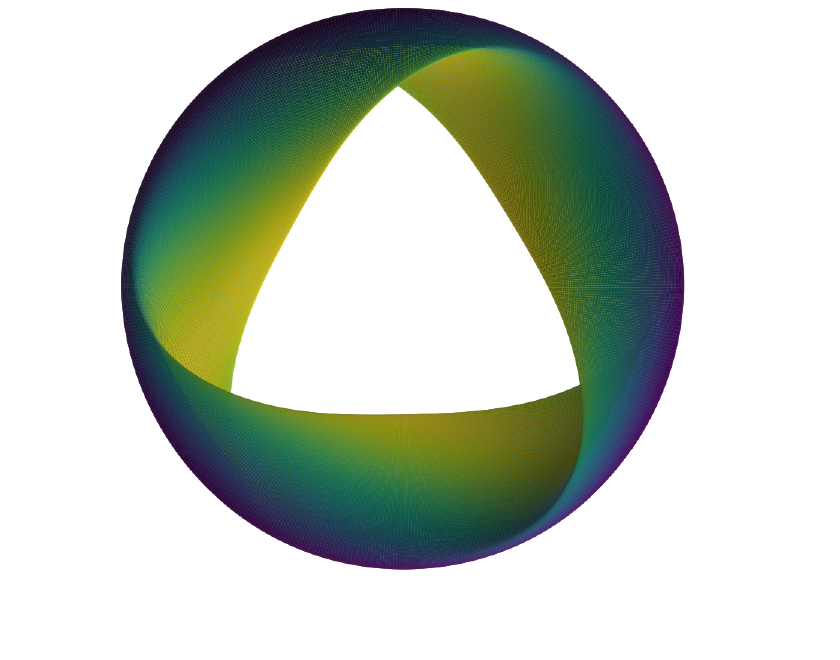}
         \label{fig:QA-Nfp3-iota-0.82-modB-top}
     \end{subfigure}
     \hfill
     \begin{subfigure}[t]{0.24\textwidth}
         \centering
         \includegraphics[width=\textwidth, trim = 0cm 1.5cm 0cm 0cm, clip]{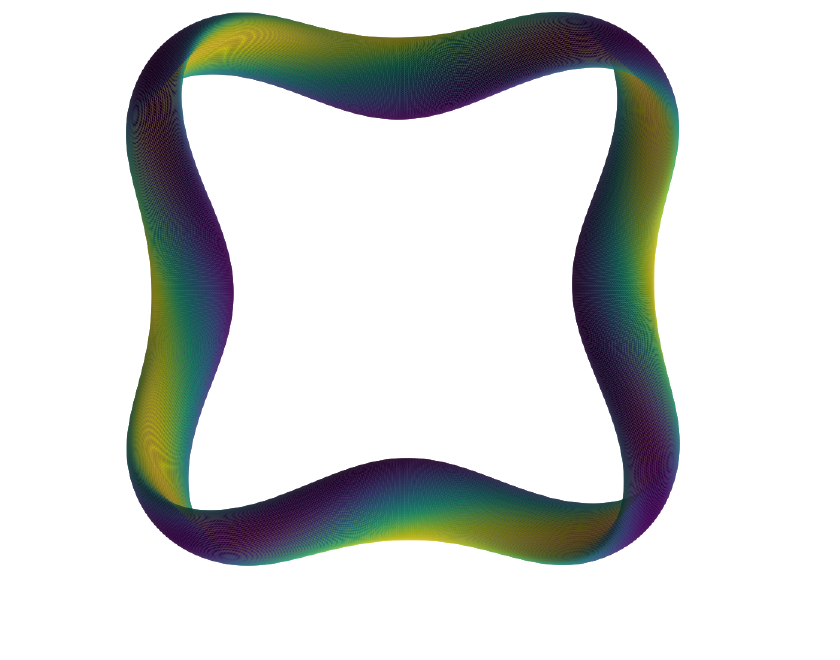}
         \label{fig:low-iota-QH-Nfp4-top}
     \end{subfigure}
     \hfill
      \begin{subfigure}[t]{0.24\textwidth}
         \centering
         \includegraphics[width=\textwidth, trim = 0cm 1.5cm 0cm 0cm, clip]{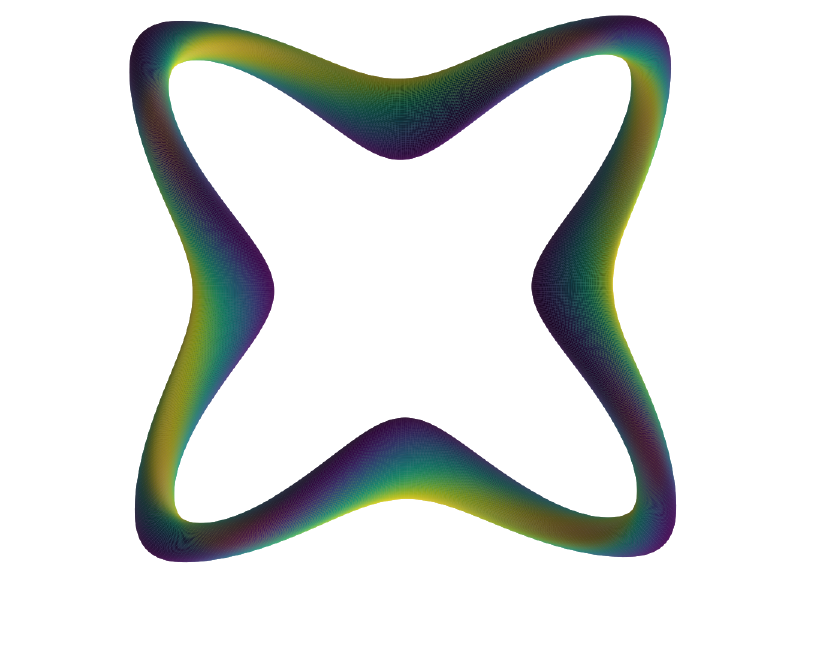}
         \label{fig:high-iota-QH-Nfp4-top}
     \end{subfigure}
     \\
     % \vspace{-7mm}
     
    \begin{subfigure}[t]{0.24\textwidth}
         \centering
         \includegraphics[width=\textwidth, trim = 0cm 3.5cm 0cm 2.2cm, clip]{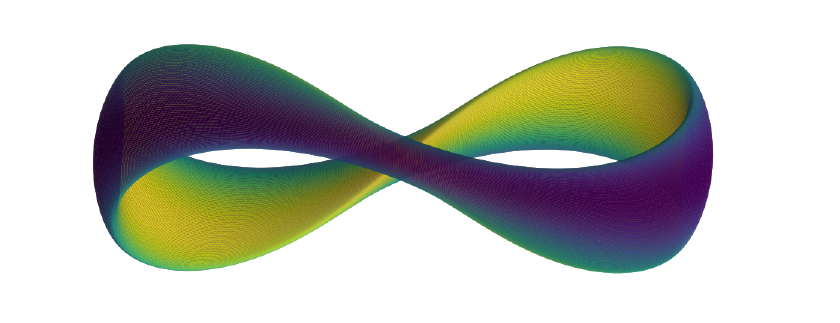}
         \label{fig:QA-Nfp2-iota-0.65-modB-side}
     \end{subfigure}
    \hfill     
    \begin{subfigure}[t]{0.24\textwidth}
         \centering
         \includegraphics[width=\textwidth, trim = 0cm 3.5cm 0cm 2.2cm, clip]{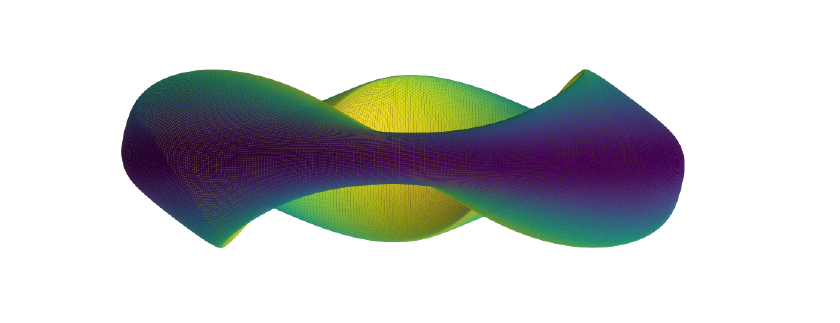}
         \label{fig:QA-Nfp3-iota-0.82-modB-side}
     \end{subfigure}
     \hfill
     \begin{subfigure}[t]{0.24\textwidth}
         \centering
         \includegraphics[width=\textwidth, trim = 0cm 3.5cm 0cm 2.2cm, clip]{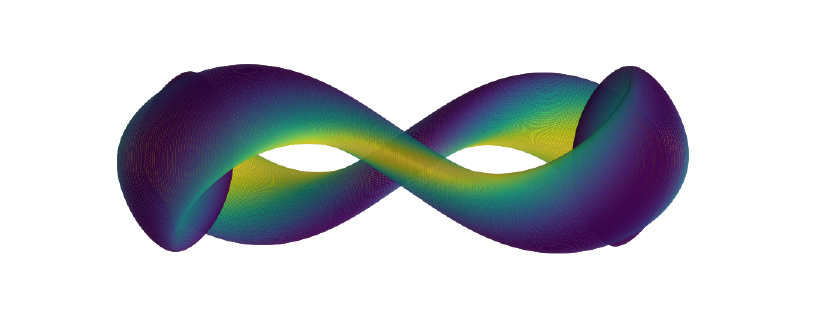}
         \label{fig:low-iota-QH-Nfp4-side}
     \end{subfigure}
     \hfill
     \begin{subfigure}[t]{0.24\textwidth}
         \centering
         \includegraphics[width=\textwidth, trim = 0cm 3.5cm 0cm 2.2cm, clip]{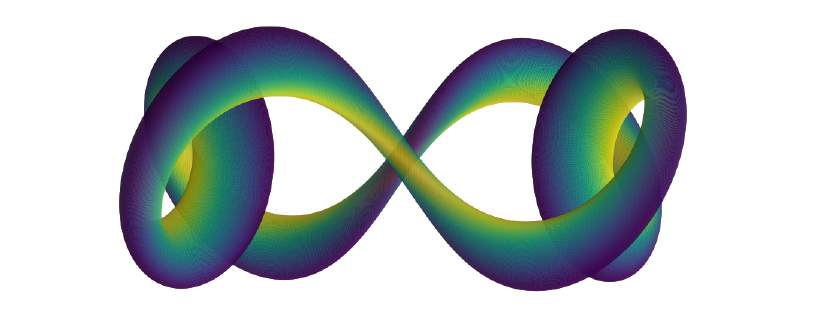}
         \label{fig:high-iota-QH-Nfp4-side}
     \end{subfigure}
     \\
     % \vspace{-12mm}
     
    \begin{subfigure}[t]{0.24\textwidth}
         \centering
         \includegraphics[width=\textwidth, trim = 0cm 0cm 0cm 1.4cm, clip]{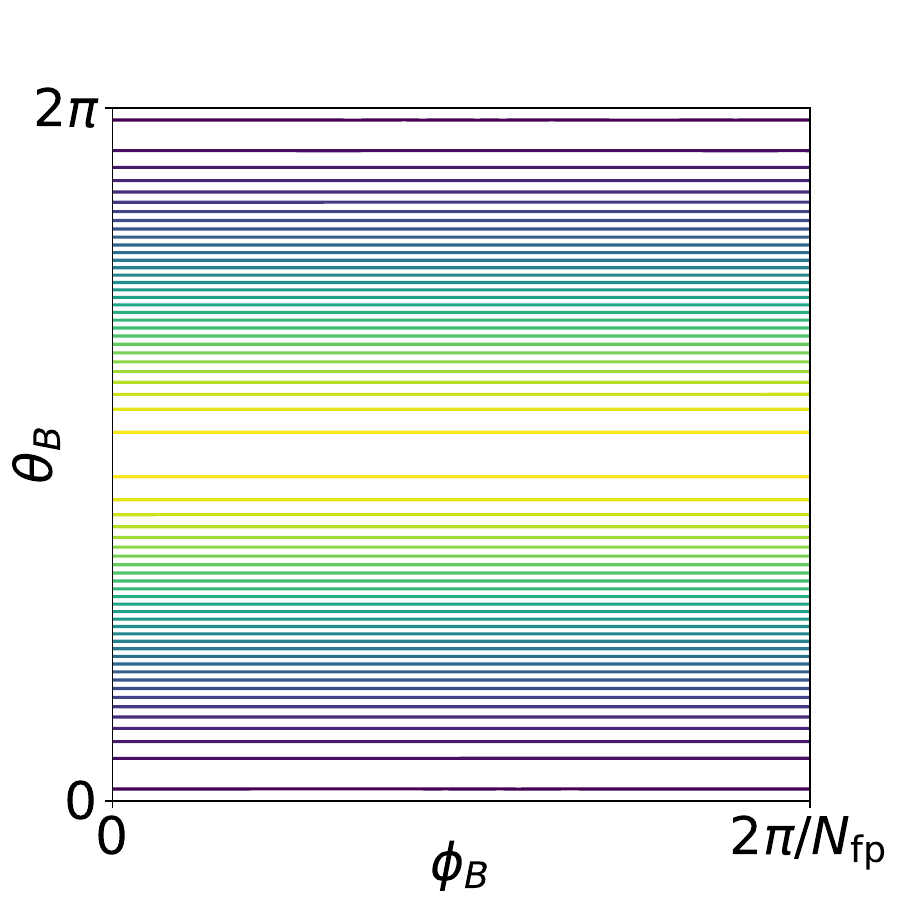}
         \caption{$\iota_e = 0.65$ $N_\mathrm{fp}=2$ QA}
     \end{subfigure}
     \hfill
     \begin{subfigure}[t]{0.24\textwidth}
         \centering
         \includegraphics[width=\textwidth, trim = 0cm 0cm 0cm 1.4cm, clip]{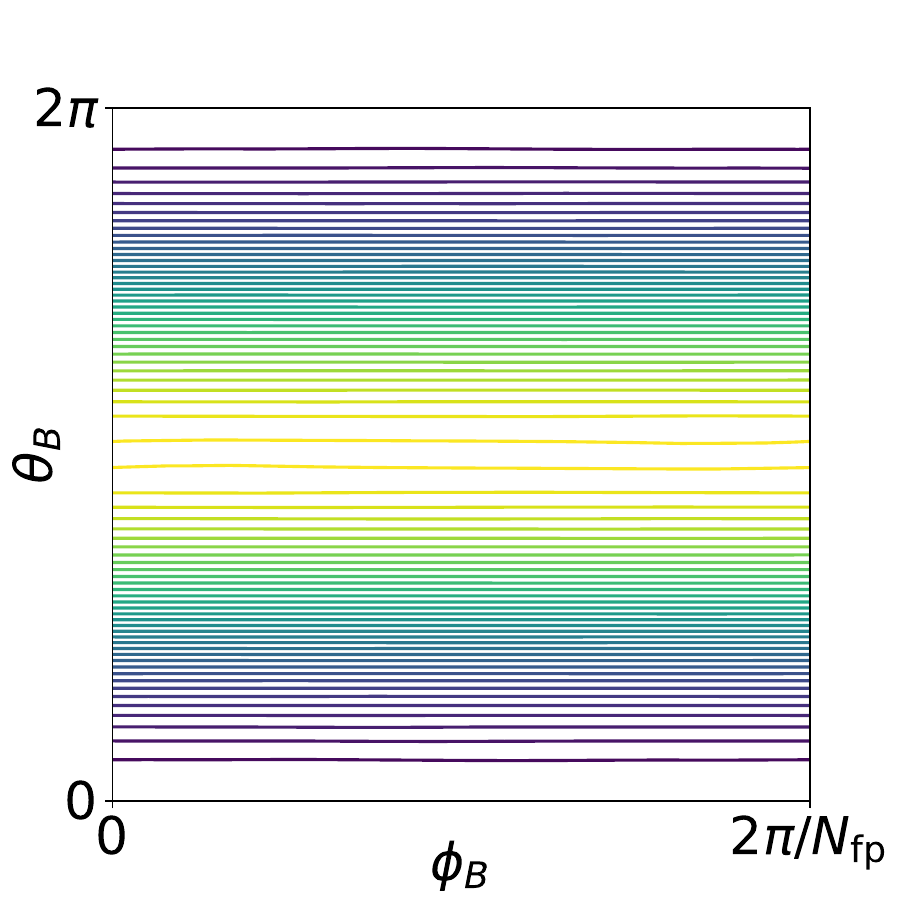}
         \caption{$\iota_e=0.82$ $N_\mathrm{fp}=3$ QA}
     \end{subfigure}
     \hfill
    \begin{subfigure}[t]{0.24\textwidth}
         \centering
         \includegraphics[width=\textwidth, trim = 0cm 0cm 0cm 1.4cm, clip]{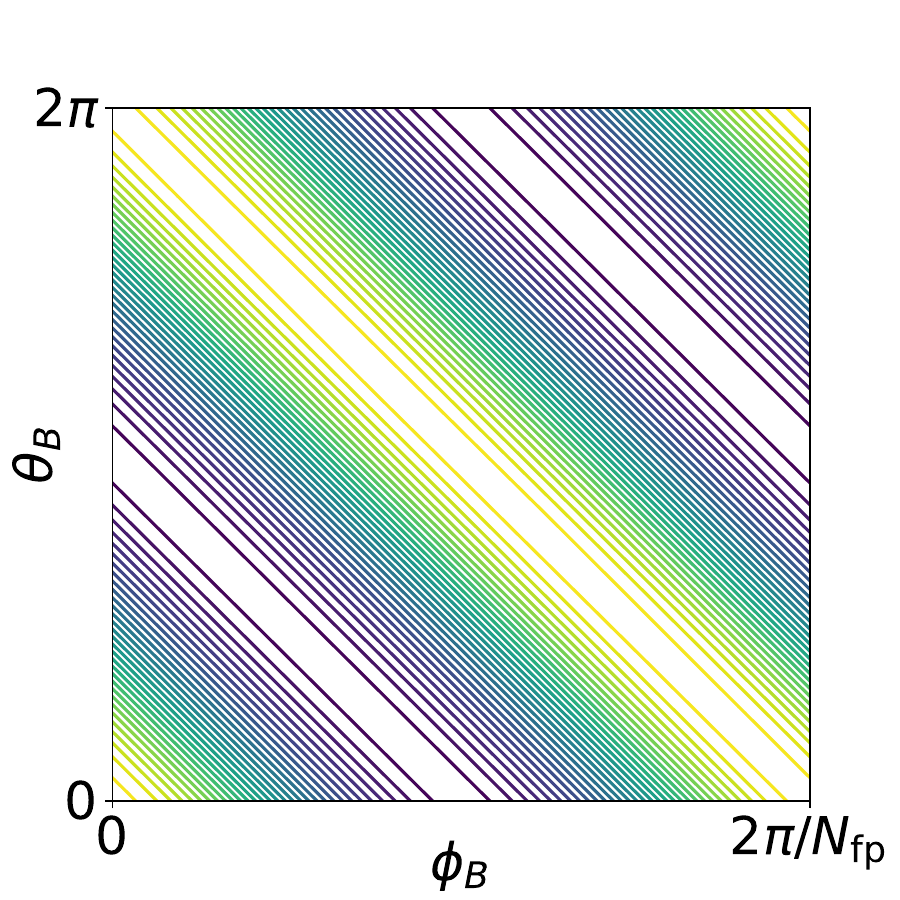}
         \caption{$\iota_e=1.12$ $N_\mathrm{fp}=4$ QH}
         \label{fig:low-iota-QH-Nfp4-Boozer}
     \end{subfigure}
    \hfill     
     \begin{subfigure}[t]{0.24\textwidth}
         \centering
         \includegraphics[width=\textwidth, trim = 0cm 0cm 0cm 1.4cm, clip]{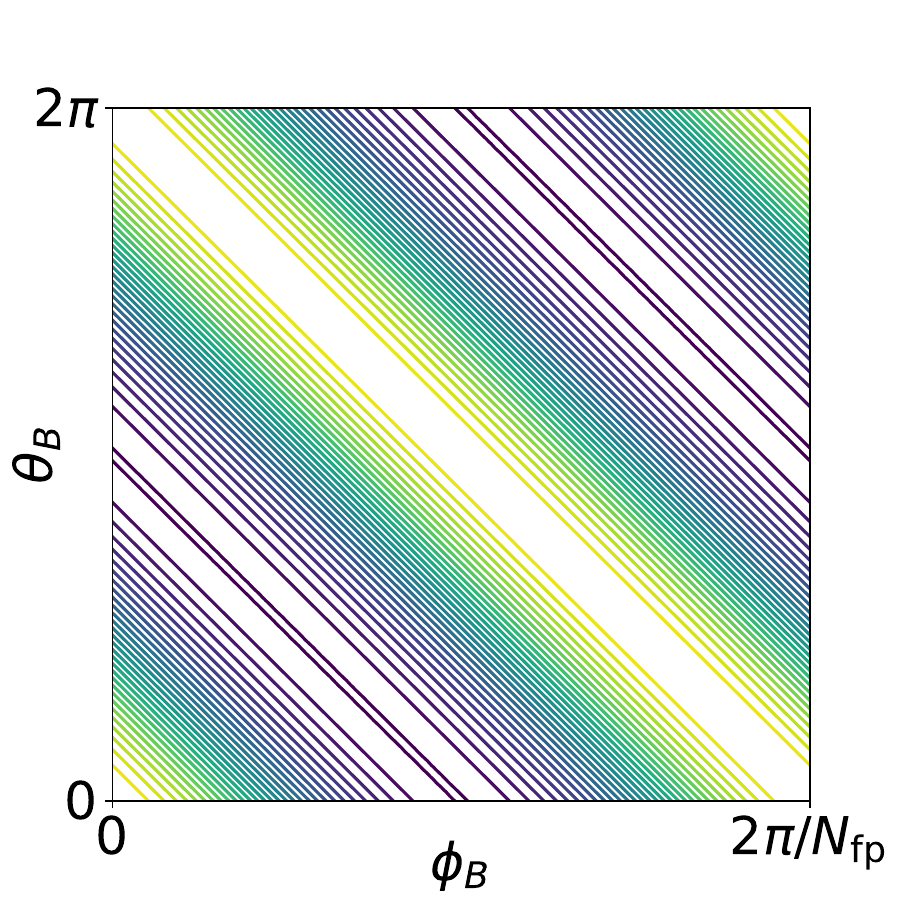}
         \caption{$\iota_e=1.97$ $N_\mathrm{fp}=4$ QH}
         \label{fig:high-iota-QH-Nfp4-Boozer}
     \end{subfigure}
        \caption{Contours of the magnetic field strength $B$ on the boundary of quasisymmetric stellarators at varying edge rotational transform $\iota_e$: top view (first row), side view (second row) and in Boozer coordinates (third row).}
         \label{fig:optimisation_highlights_iota_scan}
\end{figure}

The contours of the magnetic field strength for the QA configurations with highest rotational transform obtained ($\iota_e = 0.65$ and $\iota_e = 0.82$ for $N_\mathrm{fp}=2$ and $N_\mathrm{fp}=3$, respectively) are shown in figure~\ref{fig:optimisation_highlights_iota_scan}, alongside the contours for the QH with the lowest and highest rotational transform values obtained ($\iota_e$=1.12 and $\iota_e$=1.97). The contours of the field strength in Boozer coordinates are straight to the naked eye, attesting to the high level of QS.

\begin{figure}
     \centering
     \begin{subfigure}[t]{0.49\textwidth}
         \centering
            \includegraphics[width=\textwidth, trim = 0.4cm 0cm 0.4cm 0cm, clip]{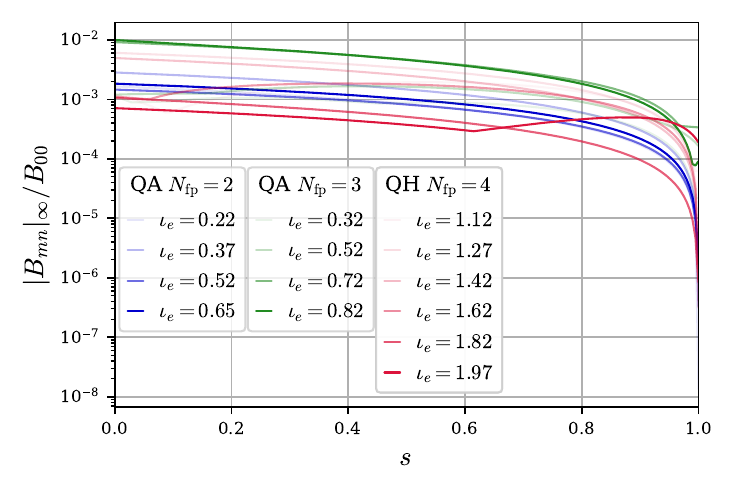}
         % \caption{Quasisymmetry error}
         % \label{fig:QA-Nfp2-iota-scan_axis_side-bmnc}
     \end{subfigure}
     \hfill
     \begin{subfigure}[t]{0.49\textwidth}
         \centering
         \includegraphics[width=\textwidth, trim = 0.4cm 0cm 0.4cm 0cm, clip]{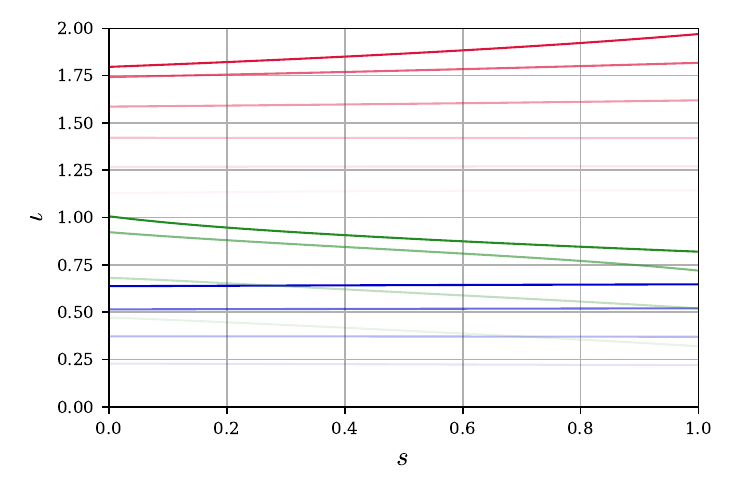}
         % \caption{Rotational transform}
         % \label{fig:QA-Nfp2-iota-scan_axis_iota}
     \end{subfigure}
        \caption{Volume properties of stellarators optimised for quasisymmetry on the boundary, with varying edge rotational transform $\iota_e$. The QS error (left plot) and rotational transform (right plot) profiles were obtained using the DESC code. The QS error reaches record levels on the boundary ($s=1$), and generally remains relatively low in the core. The rotational transform profiles are flat for all configurations except for $N_\mathrm{fp}=3$ QA and the high $\iota_e=1.97$ QH.}
\label{fig:summary_iota_scan}
\end{figure}

The quasisymmetry error in the volume and the rotational transform profile for all configurations obtained in the rotational transform scan are shown in figure~\ref{fig:summary_iota_scan}. Further properties of the configurations are listed in table~\ref{tab:iota}. The QS error is again very low at the edge in most configurations, down to $\abs{B_{mn}}_\infty/B_{00} \sim 10^{-8}$. The QS error in the core also remains low, with $\abs{B_{mn}}_\infty/B_{00} \lesssim 10^{-3}$ for most QA configurations and the high $\iota$ QH, while $\abs{B_{mn}}_\infty/B_{00} \lesssim 10^{-2}$ for the high $\iota$ $N_\mathrm{fp}=3$ QA and lower $\iota$ QH. The magnetic shear is again found to be small for the $N_\mathrm{fp}=2$ QA, and for the $N_\mathrm{fp}=4$ QH configurations at lower $\iota_e \lesssim 1.75$. Substantial magnetic shear is found for the $N_\mathrm{fp}=3$ QA configurations, with a nearly constant difference between the transform in the core and in the edge, $\Delta\iota=\iota(s=1)-\iota(s=0)\approx 0.2$. Noticeable magnetic shear is also found for the highest $\iota_e=1.97$ QH configuration.

\begin{figure}
     \centering
     \begin{subfigure}[t]{0.49\textwidth}
         \centering
         \includegraphics[width=\textwidth, trim = 0.37cm 0cm 0.4cm 0cm, clip]{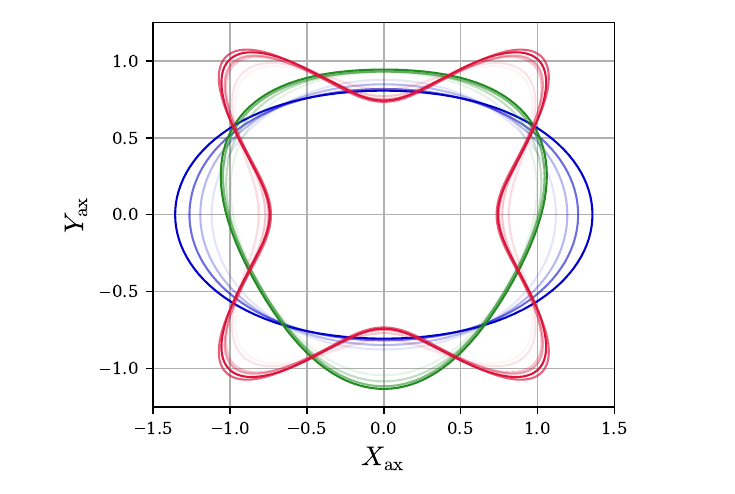}
		 % \caption{Top view}
     \end{subfigure}
     \hfill
     \begin{subfigure}[t]{0.49\textwidth}
         \centering
         \includegraphics[width=\textwidth, trim = 0.37cm 0cm 0.4cm 0cm, clip]{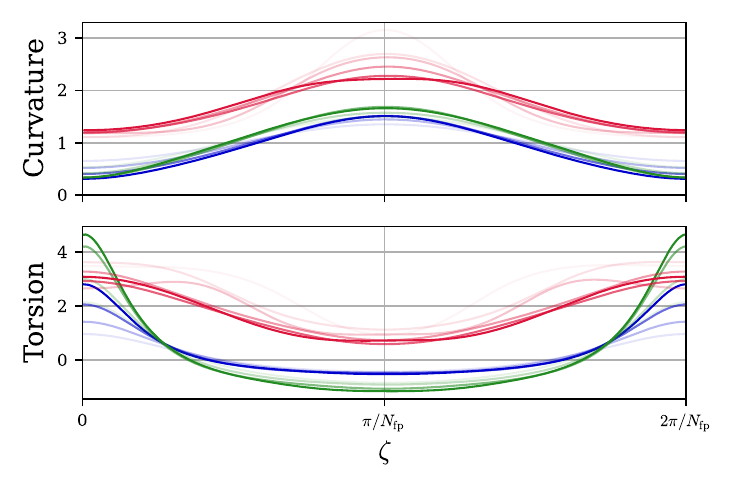}
		 % \caption{Side view}
     \end{subfigure}
        \caption{Magnetic axes top view (left plot), curvature and torsion (right plots), of the stellarators optimised for quasisymmetry on the boundary with varying edge rotational transform $\iota_e$. The DESC solution is used to obtain the magnetic axis. Regions of reduced curvature develop at high $\iota$ for the QA, and at low $\iota$ for the QH configurations, as predicted by \cite{rodriguez_phases_2022}. The legend for the various colours and shadings is given in figure~\ref{fig:summary_iota_scan}.}
        \label{fig:axis_iota_scan}
\end{figure}

The magnetic axes of the configurations at varying rotational transform are shown in figure~\ref{fig:axis_iota_scan}. For the QA configurations, as the rotational transform is increased, the axis develops regions of low curvature, as may also be seen from the straightening of the axis, and of high torsion. The straightening of the magnetic axis with increasing $\iota$ is in agreement with a near-axis model of QS \citep{rodriguez_phases_2022} that showed how higher values of $\iota$ move QA configurations towards a phase transition where the configuration would be quasi-isodynamic, which requires the axis to have regions of vanishing curvature. For the QH configurations, the phase transition is approached as $\iota$ is decreased, corresponding again to a straightening of the magnetic axis at lower $\iota$ in figure~\ref{fig:axis_iota_scan}, though it is less pronounced than for the QA cases.

\begin{table}
    \centering
    \begin{tabular}{c|c|c|c|c|c|c|c}
        & $\iota_e$ & $f_\mathrm{QS}^\star$ & $\left(\frac{\abs{B_\mathrm{mn}}_\infty}{B_{00}}\right)_{e,\mathrm{SPEC}}$ & $\left(\frac{\abs{B_\mathrm{mn}}_\infty}{B_{00}}\right)_{e,\mathrm{DESC}}$ & $\left\langle\frac{\abs{B_\mathrm{mn}}_\infty}{B_{00}}\right\rangle$ & $\Delta\iota$ & $\kappa_\mathrm{min} R_{00}$ \\\hline
\multirow{4}{6em}{QA $N_\mathrm{fp}=2$} & $0.22$ & $2\cdot10^{-7}$ & $8\cdot10^{-9}$ & $1\cdot10^{-8}$ & $8\cdot10^{-4}$ & $-8\cdot10^{-3}$ & 0.65\\
                                        & $0.37$ & $4\cdot10^{-6}$ & $3\cdot10^{-7}$ & $3\cdot10^{-7}$ & $1\cdot10^{-3}$ & $-2\cdot10^{-3}$ & 0.52\\
                                        & $0.52$ & $4\cdot10^{-5}$ & $2\cdot10^{-6}$ & $2\cdot10^{-6}$ & $8\cdot10^{-4}$ & $+5\cdot10^{-3}$ & 0.40\\
                                        & $0.65$ & $2\cdot10^{-5}$ & $2\cdot10^{-6}$ & $2\cdot10^{-6}$ & $1\cdot10^{-3}$ & $+1\cdot10^{-2}$ & 0.31\\\hline
\multirow{4}{6em}{QA $N_\mathrm{fp}=3$} & $0.32$ & $2\cdot10^{-4}$ & $3\cdot10^{-6}$ & $2\cdot10^{-5}$ & $8\cdot10^{-4}$ & $-0.152$         & 0.53\\
                                        & $0.52$ & $4\cdot10^{-4}$ & $1\cdot10^{-5}$ & $2\cdot10^{-4}$ & $1\cdot10^{-3}$ & $-0.161$         & 0.42\\
                                        & $0.72$ & $3\cdot10^{-4}$ & $7\cdot10^{-6}$ & $3\cdot10^{-4}$ & $5\cdot10^{-3}$ & $-0.203$         & 0.34\\
                                        & $0.82$ & $1\cdot10^{-3}$ & $6\cdot10^{-5}$ & $9\cdot10^{-5}$ & $4\cdot10^{-3}$ & $-0.187$         & 0.33\\\hline
\multirow{6}{6em}{QH $N_\mathrm{fp}=4$} & $1.12$ & $1\cdot10^{-3}$ & $6\cdot10^{-5}$ & $6\cdot10^{-5}$ & $1\cdot10^{-3}$ & $+0.015$         & 1.11\\
                                        & $1.27$ & $8\cdot10^{-5}$ & $2\cdot10^{-6}$ & $2\cdot10^{-6}$ & $3\cdot10^{-3}$ & $+3\cdot10^{-3}$ & 1.11\\
                                        & $1.42$ & $7\cdot10^{-6}$ & $3\cdot10^{-7}$ & $3\cdot10^{-7}$ & $3\cdot10^{-3}$ & $-2\cdot10^{-3}$ & 1.18\\
                                        & $1.62$ & $2\cdot10^{-6}$ & $1\cdot10^{-7}$ & $1\cdot10^{-6}$ & $1\cdot10^{-3}$ & $+0.033$         & 1.19\\
                                        & $1.82$ & $9\cdot10^{-6}$ & $3\cdot10^{-7}$ & $8\cdot10^{-7}$ & $5\cdot10^{-4}$ & $+0.075$         & 1.20\\
                                        & $1.97$ & $8\cdot10^{-6}$ & $1\cdot10^{-6}$ & $2\cdot10^{-4}$ & $5\cdot10^{-4}$ & $+0.173$         & 1.24
    \end{tabular}
    \caption{Properties of stellarators optimised for quasisymmetry at varying edge rotational transform $\iota_e$: QS figure of merit $f_\mathrm{QS}^\star$, maximum symmetry-breaking mode on boundary from vacuum solution (SPEC) and MHS solution (DESC), volume-averaged QS error in MHS solution, difference between $\iota$ on axis and at the edge $\Delta \iota = \iota(s=1)-\iota(s=0)$, and minimum value of axis curvature $\kappa_\mathrm{min}$.}
    \label{tab:iota}
\end{table}

%%%%%%%%%%%%%%%%%%%%%%%%%%%%%%%%%%%%%%%%%%%%%%%%%%%%%%%%%%%%%%%%%%%%%%%

\section{Optimisation refinement for volume properties} \label{sec:Auxiliary_opt}

We here use the SIMSOPT \citep{landreman_simsopt_2021} optimisation suite in conjunction with the VMEC code \cite{hirshman_three-dimensional_1986} to refine the optimisation of two configurations with good QS on the boundary. We first demonstrate improvement of QS in the volume in \S\ref{sec:postproc_QA_volume_opt}, and then show how a $N_\mathrm{fp}=3$ QA can be made integrable in \S\ref{sec:postproc_integrability_opt}. We note that the VMEC code struggles to obtain a converged equilibrium for the QH configurations at high resolutions, such that only QA configurations were considered. 

\subsection{Volume quasisymmetry} \label{sec:postproc_QA_volume_opt}

We here optimise for quasisymmetry in the volume of the $N_\mathrm{fp}=2$ QA configuration with aspect ratio $A=6$ and high rotational transform $\iota = 0.65$. The optimisation follows the procedure of \cite{landreman_magnetic_2022}. Because of the large number of harmonics employed in the adjoint optimisation ($n_\mathrm{max} = 11$ and $m_\mathrm{max} = 7$ for the $\iota_T = 0.65$ case under consideration), only a small number of steps in the SIMSOPT optimisation are computationally feasible due to the use of finite-differences to evaluate the objective function's derivative.

\begin{figure}
     \centering
     \begin{subfigure}[t]{0.49\textwidth}
         \centering
        \includegraphics[width=\textwidth]{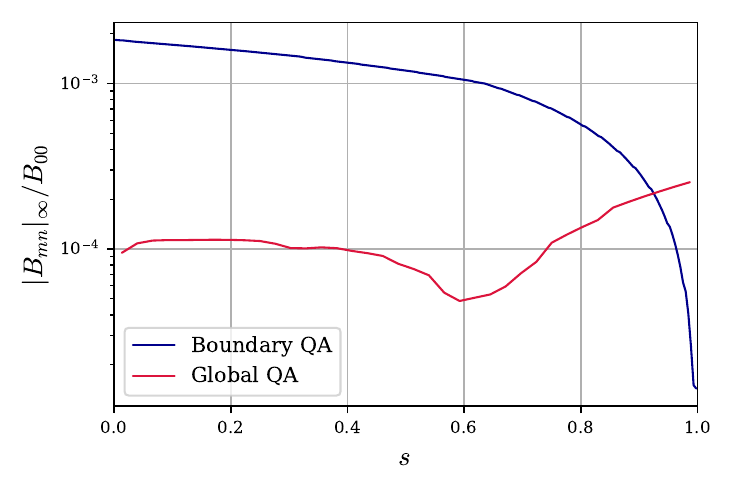}
		% \caption{Quasisymmetry error}
         \label{fig:global_opt_Nfp2_QA_bmnc}
     \end{subfigure}
     \hfill
     \begin{subfigure}[t]{0.49\textwidth}
         \centering
         \includegraphics[width=\textwidth]{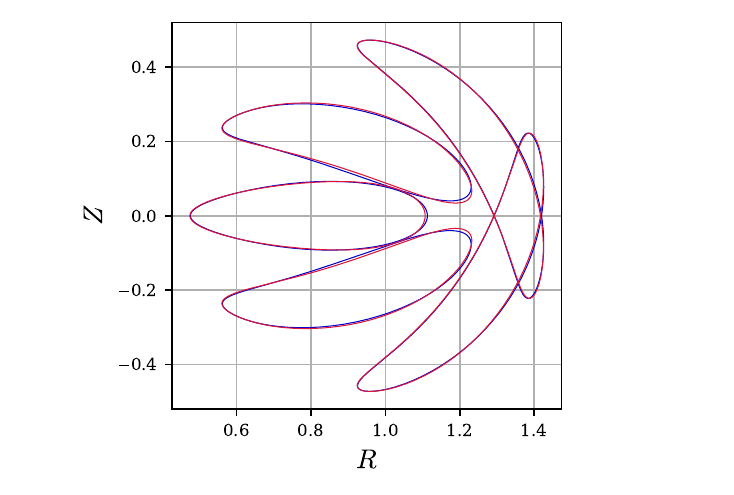}
		 %\caption{Toroidal cross-sections at $\zeta / (2\pi/5N_\mathrm{fp}) \in \{0,1,2,3,4\}$}
		 % \caption{Boundary cross-sections at varying $\zeta$}
         \label{fig:QA-Nfp2-AR-scan_cross_sections}
     \end{subfigure}
        \caption{Two field period QA with large edge rotational transform $\iota_e = 0.65$ before (Initial) and after (Optimised) auxiliary optimisation for quasisymmetry in the volume. The QS error (left plot), evaluated using the VMEC code, shows an improvement in the global QS level, at the cost of higher QS error on the boundary. The optimisation did not require a large modification to the stellarator shape, as attested by the cross-sections of the boundary (right plot), shown for $\zeta/(2\pi/N_\mathrm{fp}) \in \{0,1,2,3,4\}$.}
        \label{fig:global_opt_Nfp2_QA}
\end{figure}

After only $15$ iterations of the optimiser, the quasisymmetry in the volume is reduced by approximately an order of magnitude, although the quasisymmetry on the boundary is degraded, as shown in figure~\ref{fig:global_opt_Nfp2_QA}. The quasisymmetry error is now highest on the boundary, similar to previous optimisation results \citep[e.g.][]{landreman_magnetic_2022}. The improvement of the quasisymmetry in the volume does not require large changes in the boundary shape, as demonstrated by the small changes to the boundary cross-sections. This further supports the hypothesis that the stellarators optimised for boundary QS are close to solutions with good volume QS. Finally, we note that the already small magnetic shear is further reduced by the global QA optimisation, decreasing from $\Delta\iota = \iota(s=1)-\iota(s=0)=0.012$ to $\Delta\iota = 0.002$. 

\subsection{Flux surface nestedness}\label{sec:postproc_integrability_opt}

Generally, the quasiaxisymmetric configurations with $N_\mathrm{fp}=3$ are not integrable, i.e. they do not possess nested magnetic flux surfaces, as the magnetic shear is large enough to cause the $\iota$ profile to cross low-order rational values. Depending on the configuration at hand, this leads to magnetic island chains, or stochastic regions. In contrast, the QH and $N_\mathrm{fp}=2$ QA configurations all have nested flux surfaces, as they have small magnetic shear and the targeted $\iota$ was chosen so as to avoid low-order rationals.

Even for the non-integrable $N_\mathrm{fp}=3$ QA configurations, the integrability may be improved in a subsequent optimisation, as shown here for the $N_\mathrm{fp}=3$ QA configuration with high edge rotational transform $\iota_e=0.72$ and aspect ratio $A=6$. The integrability is targeted by minimising the Greene's Residue \citep{greene_method_1979} on a set of rational values of $\iota$ seen to cause islands in the Poincar\'e plot, as in \cite{landreman_stellarator_2021}. An objective targeting quasisymmetry in the volume is further included.

The Greene's residue is targeted for the  $\iota = 3/4$ and $\iota = 6/7$ resonances. Although the boundary coefficients go up to $m_\mathrm{max}=n_\mathrm{max}=10$, the optimisation space is limited to a small number of coefficients in the boundary representation \eqref{eqs:boundary_rep}, up to $m = 2$ and $n=4$ to make the optimisation computationally tractable.

\begin{figure}
     \centering
     \begin{subfigure}[t]{0.49\textwidth}
         \centering
         \includegraphics[width=\textwidth]{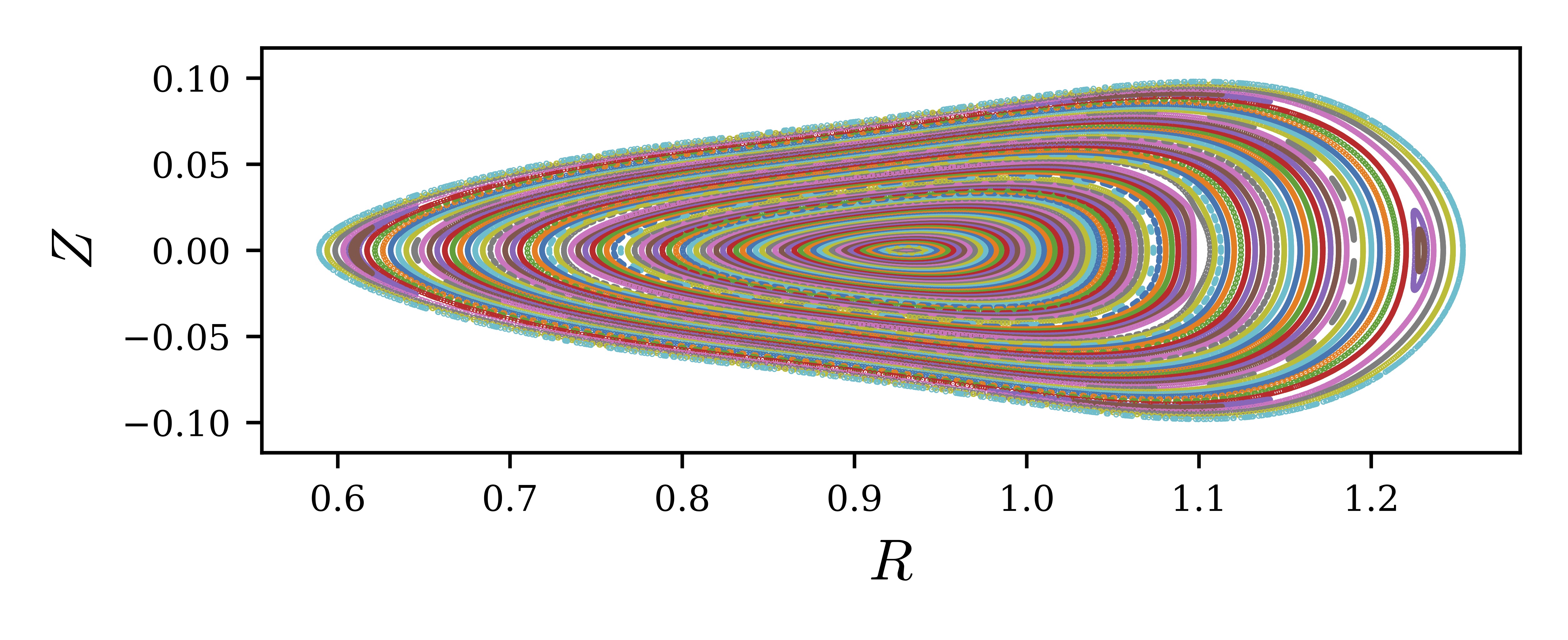}
		 \caption{Poincar\'e before integrability optimisation}
         \label{fig:QA-Nfp3-integrability-opt_poincare_pre-opt}
     \end{subfigure}
     \hfill
     \begin{subfigure}[t]{0.49\textwidth}
         \centering
         \includegraphics[width=\textwidth]{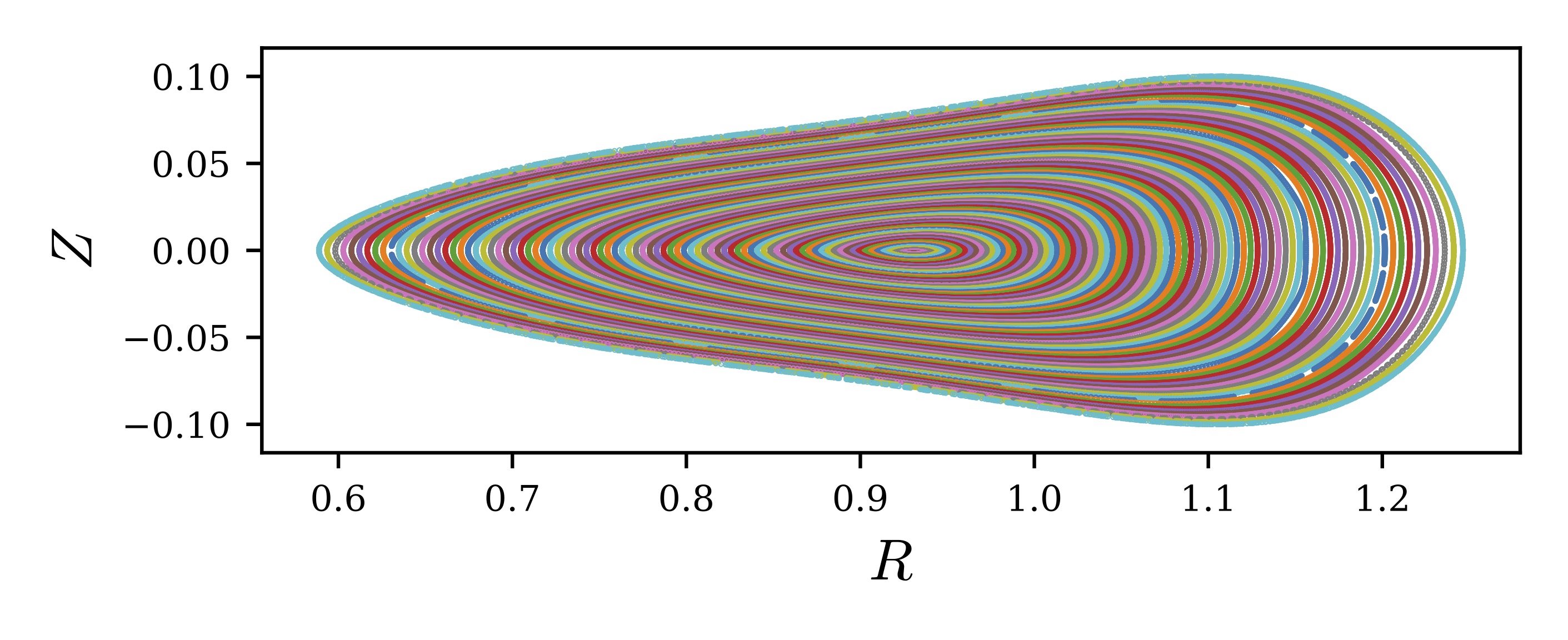}
		 \caption{Poincar\'e after integrability optimisation}
         \label{fig:QA-Nfp3-integrability-opt_poincare_post-opt}
     \end{subfigure}
     \\
     \begin{subfigure}[t]{0.49\textwidth}
         \centering
         \includegraphics[width=\textwidth]{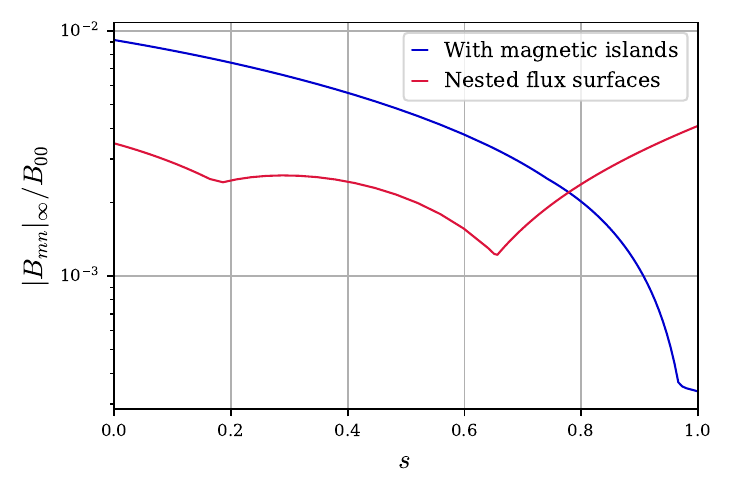}
		 \caption{Quasisymmetry error}
         \label{fig:QA-Nfp3-integrability-opt_bmnc}
     \end{subfigure}
     \hfill
     \begin{subfigure}[t]{0.49\textwidth}
         \centering
         \includegraphics[width=\textwidth]{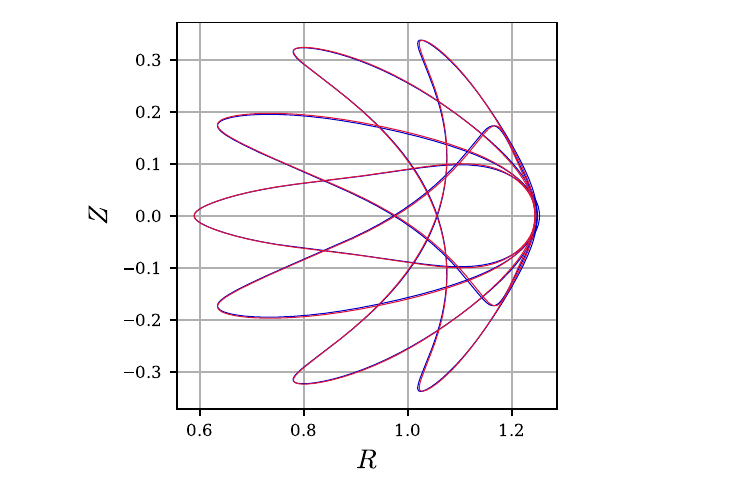}
         \caption{Boundary cross-sections at varying $\zeta$}
         \label{fig:QA-Nfp3-integrability-opt_cross-section}
     \end{subfigure}
\caption{Optimisation for integrability and volume quasisymmetry of $N_\mathrm{fp}=3$ QA with high edge rotational transform $\iota_e=0.72$. The integrability optimisation leads to the disappearance of the magnetic island chains, such that the final configuration has optimised configuration. Furthermore, the volume QS error (evaluated using the VMEC code) is also reduced. The required change in the stellarator shape is minimal, as attested by the cross-sections of the boundary at $\zeta/(2\pi/N_\mathrm{fp}) \in \{0,1,2,3,4\}$.}
         \label{fig:QA-Nfp3-integrability-opt_poincare}
\end{figure}

Before the integrability optimisation, the Poincar\'e plots clearly reveal the presence of multiple magnetic island chains, see figure~\ref{fig:QA-Nfp3-integrability-opt_poincare_pre-opt}. These are removed by the optimisation, as shown in figure~\ref{fig:QA-Nfp3-integrability-opt_poincare_post-opt}. Furthermore, the quasisymmetry error in the core could be reduced, though the quasisymmetry error on the boundary is degraded, see figure~\ref{fig:QA-Nfp3-integrability-opt_bmnc}. The changes in the quasisymmetry and integrability do not require a substantial modification to the boundary shape, as demonstrated by the cross-sections in Fig~\ref{fig:QA-Nfp3-integrability-opt_cross-section}.

\section{Conclusions} \label{sec:conclusions}

We leveraged the computational efficiency of adjoint methods to optimise for quasisymmetry on the boundary of stellarator vacuum fields, reaching record levels of QS on the boundary. For a $N_\mathrm{fp}=2$ quasi-axisymmetric configuration with $A=6$ and $\iota_e$ similar to \citep{landreman_magnetic_2022}, where a volume symmetry-breaking mode amplitude of $\abs{B_{mn}}_\infty/B_{00} \approx 3 \cdot 10^{-5}$ was achieved, we could optimise quasisymmetry to $\abs{B_{mn}}_\infty/B_{00} \approx 2 \cdot 10^{-10}$ on the boundary. The configurations optimised for boundary QS appear close to solutions with precise global QS, such that the boundary QS optimisation may be followed by a global QS optimisation, the latter requiring only small changes to the boundary, see \S\ref{sec:postproc_QA_volume_opt}.

We were also able to obtain quasiaxisymmetric configurations at $N_\mathrm{fp}=3$, which possess substantial magnetic shear compared to the QH and $N_\mathrm{fp}=2$ QA configurations \citep[see also][]{landreman_magnetic_2022}. An increased magnetic shear could be beneficial, e.g. to reduce the turbulent transport. However, the larger magnetic shear generally leads to the rotational transform crossing resonances, causing magnetic island chains and stochastic regions. We showed in \S\ref{sec:postproc_integrability_opt} how these may be removed in an auxiliary optimisation, restoring flux surface nestedness.

We studied in \S\ref{sec:aspect_ratio_scan} the range of aspect ratio values for which good QS could be obtained, leading to designs with small aspect ratios similar to those of tokamaks, e.g. an $A=2.6$ QA with $N_\mathrm{fp}=2$, and QH configurations with $A=3.6$ for $N_\mathrm{fp}=3$, and $A=3$ for $N_\mathrm{fp}=4$. Reaching such low aspect ratios could prove crucial for a stellarator fusion power plant, as a large aspect ratio might require a prohibitively large major radius for a prescribed volume target. However, configurations with tight aspect ratios have degraded levels of QS and generally have a large amount of shaping, for which economical coils could be precluded.

We further investigaged in \S\ref{sec:iota_scan} the range of $\iota$ values compatible with good QS on the boundary. Quasiaxisymmetric configurations could be found up to $\iota \sim 0.7-0.8$, while the quasihelical configurations at $N_\mathrm{fp}=4$ had a range of $\iota$ between $\sim 1$ and $2$. The value of $\iota$ is crucial in many ways, e.g. to avoid prompt losses of energetic particles \citep{paul_energetic_2022}, as the orbit width scales inversely with $\abs{\iota-N/M}$. For the QA configurations, good QS is most easily obtained at low rotational transform values, as these are closer to the axisymmetric case. For the QH configurations, good QS is most readily obtained at intermediate values of $\iota_e \sim 1.5$, though the high $\iota_e \sim 2$ configurations might prove interesting due to their increased magnetic shear.

We note that although the inclusion of thermal pressure and plasma current can lead to significant changes to the QS solutions \citep[e.g.][]{landreman_optimization_2022}, vacuum configurations can provide useful guidance for finite-pressure equilibria \citep{boozer_curl-free_2019}. Furthermore, although the present study was mostly limited to optimisation of QS on the boundary, this might be sufficient to guarantee good confinement by effectively creating an edge transport barrier (for the guiding centre trajectories). Moreover, reducing the confinement in the plasma core may be desirable to avoid impurity and ash accumulation. If global QS is desired however, the configurations presented generally possess good levels of QS even in the core, which may be further improved by a subsequent global QS optimisation, see \S\ref{sec:postproc_QA_volume_opt}.

In this study, we focused on the compatibility of quasisymmetry with aspect ratio and rotational transform targets. Future work ought to consider the trade-offs between quasisymmetry and other targets typically considered in the design of a stellarator, such as MHD stability, turbulent transport optimisation, or engineering feasibility. Furthermore, adjoint-based fixed-boundary optimisation could be combined with coil optimisation for an efficient derivative-based single-stage approach simultaneously optimising the plasma and the coils, as studied by \cite{jorge_single-stage_2023}.

All configurations presented in this study may be found in \cite{nies_data_2024}.

\begin{acknowledgments}
This work was supported by U.S. DOE DE-AC02-09CH11466, DE-SC0016072, DE-AC02–76CH03073, DE-SC0024548, and by the Simons Foundation/SFARI (560651, AB). R.N. thanks Eduardo Rodriguez, Stefan Buller and Matt Landreman for helpful discussions.
\end{acknowledgments}

\appendix

\section{Normalised quasisymmetry objective} \label{app:normalised_qs_obj}

We introduce two normalisations in our quasisymmetry objective, compared to \cite{nies_adjoint_2022}. As a reminder, our objective is based on the definition of quasisymmetry as 
\begin{equation}
    \frac{\mathbf{B}\cdot\nabla\psi\times\nabla B}{\mathbf{B}\cdot\nabla B} = -\frac{MG + NI}{N-\iota M},
\end{equation}
with the magnetic field $\mathbf{B}$, the toroidal flux $\psi$, and the enclosed poloidal and toroidal currents, $G$ and $I$, respectively. We further consider a vacuum magnetic field
\begin{equation}
    \mathbf{B}=G\nabla(\zeta+\omega) \equiv G \mathbf{\breve{B}}, \label{eq:vacuum_field}
\end{equation}
with $I=0$, $\omega$ a single-valued scalar function, and $\mathcal{S}$ the stellarator boundary. The quasisymmetry objective in \cite{nies_adjoint_2022} was defined as 
\begin{equation}
    f_\mathrm{QS} = \int_{\mathcal{S}} \mathrm{d}S\; \frac{1}{2}v_\mathrm{QS}^2, \qquad \text{with} \qquad v_\mathrm{QS} =  \mathbf{\breve{B}}\cdot\nabla \breve{B}  -  \mathbf{\breve{B}}\times\mathbf{\breve{g}}_\psi\cdot\nabla \breve{B}  \left( \iota - N/M\right),
\end{equation}
where $\mathbf{\breve{B}}=\mathbf{B}/G$, and $\mathbf{\breve{g}}_\psi$ is a vector defined on the boundary that reduces to $\nabla\psi/G$ in the limit of an integrable magnetic field, see \cite{nies_adjoint_2022}. In this study, we consider the objective
\begin{equation}
    f_\mathrm{QS}^\star = \sqrt{\int_{\mathcal{S}} \mathrm{d}S\; w_\mathrm{QS}^2} \qquad \text{with} \qquad w_\mathrm{QS} = v_\mathrm{QS} / \breve{B}^2. \label{eq:def_fQS_star}
\end{equation}
The normalisation to $w_\mathrm{QS}$ makes the objective function dimensionless (as $w_\mathrm{QS}\sim 1/L$), and should lend itself better to QP optimisation, as $f_\mathrm{QS}$ is dominated by contributions from high-field regions and QP configurations typically exhibits significant variation of $B$ on the surface. Finally, the square root is motivated by the fact that the Fourier expansion of $w_\mathrm{QS}$ in Boozer coordinates \citep{rodriguez_measures_2022} is
\begin{equation}
    w_\mathrm{QS} = -i \frac{1}{G} \sum_{n,m} \left(n-m N/M\right) B_{mn} e^{i(m\theta_B-n\zeta_B)},
\end{equation}
so that we can expect our objective to scale linearly in the size of the symmetry-breaking Fourier components.

We now compute the shape derivative of the new quasisymmetry objective \eqref{eq:def_fQS_star}. First, note that
\begin{equation}
    \delta f_\mathrm{QS}^\star = \frac{1}{f_\mathrm{QS}^\star} \delta\left( \int_{\mathcal{S}} \mathrm{d}S\; \frac{1}{2} w_\mathrm{QS}^2 \right),
\end{equation}
and
\begin{align}
    & \delta\left( \int \mathrm{d}S\; \frac{1}{2} w_\mathrm{QS}^2 \right)  = \int_{\mathcal{S}} \mathrm{d}S\; \left[ w_\mathrm{QS} \; \delta w_\mathrm{QS} + \frac{1}{2}\left(\normpert\right) (\normvec\cdot\nabla + h)w_\mathrm{QS}^2 \right] \\
    & = \int_{\mathcal{S}} \mathrm{d}S\; \left\{ \frac{v_\mathrm{QS}}{\breve{B}^4}\delta v_\mathrm{QS} + \delta\omega \nabla_\Gamma\cdot \left( 2 \frac{v_\mathrm{QS}^2}{\breve{B}^6} \mathbf{\breve{B}} \right) + (\normpert) \left[ \frac{v_\mathrm{QS}}{\breve{B}^4} \normvec\cdot\nabla v_\mathrm{QS} - 2 \frac{v_\mathrm{QS}^2}{\breve{B}^5}\normvec\cdot\nabla\breve{B} + \frac{h}{2}\frac{v_\mathrm{QS}^2}{\breve{B}^4}  \right] \right\},
\end{align}
where we used $\delta \breve{B} = \mathbf{\breve{B}}\cdot\nabla(\delta\omega)/\breve{B}$ and partially integrated the second term. Here $h$ is the summed curvature, and $\normvec$ is a unit vector normal to the surface. We can now proceed entirely analogously to the derivation of $\delta f_\mathrm{QS}$ in (B24) of \cite{nies_adjoint_2022}, normalising by $\breve{B}^{-4}$ where $v_\mathrm{QS}$ appears, leading to
\begin{align}
     \delta & f_\mathrm{QS}^\star = \frac{1}{f_\mathrm{QS}^\star}\int_{\mathcal{S}} \mathrm{d} S\; \Bigg\{   \delta\omega \;\nabla_\Gamma\cdot \Bigg[ -\frac{v_\mathrm{QS}}{\breve{B}^4} \nabla_\Gamma \breve{B} + \frac{\mathbf{\breve{B}}}{\breve{B}} \nabla_\Gamma\cdot\left(\frac{v_\mathrm{QS}}{\breve{B}^4} \mathbf{\breve{B}}\right)  \label{eq:delta_fQS_tot} \\
     \nonumber & + \left(\iota-N/M\right) \Bigg(  \frac{v_\mathrm{QS}}{\breve{B}^4} \; \mathbf{\breve{g}}_\psi\times\nabla_\Gamma\breve{B} - \mathbf{\breve{B}}\; \nabla_\Gamma\cdot\left(\frac{1}{\breve{B}} \frac{v_\mathrm{QS}}{\breve{B}^4} \mathbf{\breve{B}}\times\mathbf{\breve{g}}_\psi\right) \Bigg) + 2\frac{v_\mathrm{QS}^2}{\breve{B}^6} \mathbf{\breve{B}}  \Bigg]\\
     \nonumber & - \delta\iota \; \frac{v_\mathrm{QS}}{\breve{B}^4} \mathbf{\breve{B}}\times\mathbf{\breve{g}}_\psi\cdot\nabla_\Gamma\breve{B}  \left[ \left(\iota-N/M\right) \frac{\nabla_\Gamma\alpha\cdot\nabla_\Gamma\zeta}{\abs{\nabla_\Gamma\alpha}^2} + 1\right]  \\
     \nonumber & - \delta\lambda \;\left(\iota-N/M\right)\;\nabla_\Gamma\cdot\left[ \frac{v_\mathrm{QS}}{\breve{B}^4}  \frac{\nabla_\Gamma\alpha}{\abs{\nabla_\Gamma\alpha}^2} \; \mathbf{\breve{B}}\times\mathbf{\breve{g}}_\psi\cdot\nabla_\Gamma\breve{B} \right] \\
     \nonumber & + (\normpert)\; \Bigg[ \left(\iota-N/M\right)\; \abs{\mathbf{\breve{g}}_\psi} \; \mathbf{\breve{B}}\times\nabla\breve{B}\cdot \left( \nabla_\Gamma \left( \frac{v_\mathrm{QS}}{\breve{B}^4} \right) - \frac{v_\mathrm{QS}}{\breve{B}^4} \frac{\nabla_\Gamma \abs{\nabla_\Gamma\alpha}}{\abs{\nabla_\Gamma\alpha}} \right)\\ 
    \nonumber & + \frac{h}{2} \frac{v_\mathrm{QS}^2}{\breve{B}^4}  - (\normvec\cdot\nabla\breve{B}) \nabla_\Gamma\cdot\left( \frac{v_\mathrm{QS}}{\breve{B}^4} \mathbf{\breve{B}} \right) - \frac{v_\mathrm{QS}}{\breve{B}^4}\;\left(\mathbf{\breve{B}}\cdot\nabla\normvec - \normvec\cdot\nabla\mathbf{\breve{B}} \right)\cdot\nabla_\Gamma\breve{B}   \\
    \nonumber & - \frac{v_\mathrm{QS}}{\breve{B}^4} \; \mathbf{\breve{B}}\times\mathbf{\breve{g}}_\psi\cdot\nabla \breve{B}  \;  \left( \iota - N/M\right) \left( \frac{\nabla_\Gamma\alpha\cdot\nabla\normvec\cdot\nabla_\Gamma\alpha}{\abs{\nabla_\Gamma \alpha}^2} -h \right) - 2 \frac{v_\mathrm{QS}^2}{\breve{B}^5}\normvec\cdot\nabla\breve{B} \Bigg] \Bigg\},
\end{align}
where $\nabla_\Gamma$ is the tangential derivative. These modifications then carry through to the adjoint equations and shape gradient, reproduced here for convenience. 

First, the adjoint $q_\alpha$ to the straight field line equation $\mathbf{B}\cdot\nabla\alpha=0$, with $\alpha = \theta - \iota\zeta + \lambda$ and single-valued $\lambda$, is given by
\begin{subequations}
\begin{align}
   \nabla_\Gamma \cdot \Big( q_\alpha \mathbf{\breve{B}} \Big) = -\frac{1}{f_\mathrm{QS}^\star} \nabla_\Gamma\cdot\left[ \nabla_\Gamma\alpha \left( \frac{v_\mathrm{QS}}{\breve{B}^4}\; \mathbf{\breve{B}}\times\mathbf{\breve{g}}_\psi\cdot\nabla \breve{B}\; \frac{\iota - N/M}{\abs{\nabla_\Gamma \alpha}^2}\right) \right] \label{eq:QS_fom_adjoint_diff_eq_qsfl}, \\
   0=\int_{\mathcal{S}} \mathrm{d} S \left\{ q_\alpha \mathbf{\breve{B}} \cdot \nabla\zeta +  \frac{v_\mathrm{QS}}{f_\mathrm{QS}^\star\breve{B}^4}\; \mathbf{\breve{B}}\times\mathbf{\breve{g}}_\psi\cdot\nabla \breve{B} \left[ \frac{\nabla_\Gamma\alpha\cdot\nabla_\Gamma \zeta}{\abs{\nabla_\Gamma \alpha}^2} (\iota-N/M) + 1 \right] \right\} \label{eq:QS_fom_adjoint_integral_eq_qsfl}.
\end{align}
\end{subequations}

Second, the adjoint $q_\omega$ to the Laplace equation $\Delta \omega = 0$ may be written as
\begin{subequations}
\begin{align}
    \Delta q_\omega & = 0   \qquad\qquad\qquad\mathrm{in}\;\mathcal{V}, \label{eq:QS_fom_adjoint_eq_qomega}\\
    \nabla q_\omega \cdot \normvec &= -\nabla_\Gamma \cdot \Bigg\{ q_\alpha \nabla_\Gamma \alpha + \frac{v_\mathrm{QS}}{f_\mathrm{QS}^\star\breve{B}^4} \;\nabla_\Gamma \breve{B} - \frac{\mathbf{\breve{B}}}{f_\mathrm{QS}^\star\breve{B}} \nabla_\Gamma \cdot \left(\frac{v_\mathrm{QS}}{\breve{B}^4} \; \mathbf{\breve{B}}\right) - 2\frac{v_\mathrm{QS}^2}{f_\mathrm{QS}^\star\breve{B}^6} \mathbf{\breve{B}} \label{eq:QS_fom_adjoint_eq_qomega_normal_BC} \\
    \nonumber & + \frac{1}{f_\mathrm{QS}^\star}\left(\iota-N/M\right) \left[ \frac{v_\mathrm{QS}}{\breve{B}^4} \; \mathbf{\breve{g}}_\psi\times\nabla_\Gamma\breve{B} - \mathbf{\breve{B}}\; \nabla_\Gamma\cdot\left(\frac{1}{\breve{B}}\frac{v_\mathrm{QS}}{\breve{B}^4} \mathbf{\breve{B}}\times\mathbf{\breve{g}}_\psi\right)  \right] \Bigg\} \quad \text{ on } \mathcal{S},
\end{align}
\end{subequations}
with $\mathcal{V}$ the volume enclosed by $\mathcal{S}$.

Finally, the shape gradient follows as
\begin{align}
    \mathcal{G}&_\mathrm{QS}^\star  =  -\frac{1}{f_\mathrm{QS}^\star} (\normvec\cdot\nabla\breve{B}) \nabla_\Gamma\cdot\left( \frac{v_\mathrm{QS}}{\breve{B}^4} \mathbf{\breve{B}} \right) -\frac{1}{f_\mathrm{QS}^\star} \frac{v_\mathrm{QS}}{\breve{B}^4}\;\left(\mathbf{\breve{B}}\cdot\nabla\normvec - \normvec\cdot\nabla\mathbf{\breve{B}} \right)\cdot\nabla_\Gamma\breve{B} \label{eq:QS_fom_shape_gradient} \\
    \nonumber & + \frac{\iota-N/M}{f_\mathrm{QS}^\star} \abs{\mathbf{\breve{g}}_\psi} \; \mathbf{\breve{B}}\times\nabla\breve{B}\cdot  \left[ \abs{\nabla_\Gamma\alpha} \nabla_\Gamma\left( \frac{v_\mathrm{QS}}{\breve{B}^4\abs{\nabla_\Gamma\alpha}} \right) + \normvec\; \frac{v_\mathrm{QS}}{\breve{B}^4}\; \left( \frac{\nabla_\Gamma\alpha\cdot\nabla\normvec\cdot\nabla_\Gamma\alpha}{\abs{\nabla_\Gamma \alpha}^2} -h \right) \right]  \\
    \nonumber & + \mathbf{\breve{B}}\cdot\nabla q_\omega + q_\alpha \left( \normvec\cdot\nabla\mathbf{\breve{B}} - \mathbf{\breve{B}}\cdot\nabla\normvec \right)\cdot\nabla_\Gamma\alpha + \frac{1}{f_\mathrm{QS}^\star}\frac{h}{2} \frac{v_\mathrm{QS}^2}{\breve{B}^4} - 2 \frac{1}{f_\mathrm{QS}^\star}\frac{v_\mathrm{QS}^2}{\breve{B}^5}\normvec\cdot\nabla\breve{B}.
\end{align}

\section{Shape gradient for aspect ratio objective} \label{app:aspectratio_sg}

The definition of the aspect ratio $A$ is given in \eqref{eq:def_aspectratio}. Note that the mean cross-sectional area may be written as
\begin{equation}
    S_\zeta = \frac{1}{2\pi} \int\mathrm{d}\zeta \int_{\mathcal{S}_\zeta} \mathrm{d}S = \frac{1}{2\pi}\int_\mathcal{V}\mathrm{d}V\; \sqrt{\frac{g_{ss}g_{\theta\theta}-g_{s\theta}^2}{g}},
\end{equation}
with the metric elements $g_{ii}$ and its determinant $g$. The shape derivative follows as \citep[see e.g.][]{walker_shapes_2015},
\begin{equation}
    \delta S_\zeta = \frac{1}{2\pi}\int_\mathcal{S}\mathrm{d}S\; (\normpert) \sqrt{\frac{g_{ss}g_{\theta\theta}-g_{s\theta}^2}{g}}.
\end{equation}
One may then obtain the shape derivative of the aspect ratio \eqref{eq:def_aspectratio} as
\begin{equation}
    \delta A = A \left( \frac{\delta \mathcal{V}}{\mathcal{V}} - \frac{3}{2}\frac{\delta S_\zeta}{S_\zeta} \right),
\end{equation}
with the enclosed volume's shape derivative \citep[see e.g.][]{walker_shapes_2015} $\delta \mathcal{V} = \int_\mathcal{S}\mathrm{d}S\; (\normpert)$.

\section{Boozer coordinate transformation for a vacuum magnetic field} \label{app:Boozer_transf}

The covariant coordinates of the magnetic field in Boozer coordinates $(\psi, \theta_B, \zeta_B)$ are given by
\begin{equation}
    \mathbf{B} = G(\psi) \nabla \zeta_B + I(\psi)\nabla \theta_B + K(\bf{r})\nabla\psi,
\end{equation}
which reduces to $\mathbf{B} = G \nabla\zeta_B$ for a vacuum magnetic field. Starting from a general set of coordinates $(s, \theta, \zeta)$, one may also write the vacuum field as
\begin{equation}
    \mathbf{B} = G\nabla(\zeta + \omega),
\end{equation}
with $\omega$ a single-valued function. The toroidal Boozer angle may thus be simply identified as
\begin{equation}
    \zeta_B = \zeta + \omega.
\end{equation}
The poloidal Boozer angle can be obtained after solving the magnetic differential equation $\mathbf{B}\cdot\nabla\alpha = 0$ for the field line label $\alpha = \theta - \iota \zeta + \lambda(\theta, \zeta)$, with the single-valued function $\lambda$. Then, employing the fact that Boozer coordinates are straight field-line coordinates ($\lambda=0$), the Boozer poloidal angle may be evaluated as
\begin{equation}
    \theta_B = \alpha + \iota \zeta_B.
\end{equation}

\bibliographystyle{jpp}
\bibliography{references_clean}

\end{document}